\documentclass[12pt,epsf]{article}

\usepackage{epsfig}

\begin{document}
\baselineskip=0.7cm
\newcommand{\EQ}{\begin{equation}}
\newcommand{\EN}{\end{equation}}
\newcommand{\EQA}{\begin{eqnarray}}
\newcommand{\EQN}{\end{eqnarray}}
\newcommand{\EQAN}{\begin{eqnarray*}}
\newcommand{\EQNN}{\end{eqnarray*}}
\newcommand{\e}{{\rm e}}
\newcommand{\Sp}{{\rm Sp}}
\newcommand{\Tr}{{\rm Tr}}
\newcommand{\lpartial}{\buildrel \leftarrow \over \partial}
\newcommand{\rpartial}{\buildrel \rightarrow \over \partial}
\newcommand{\nn}{\nonumber}
\def\adss{{\rm AdS}_5\times S^{5}}
\def\s5{S^{5}}
\def\ft#1#2{{\textstyle{{\scriptstyle #1}\over {\scriptstyle #2}}}}
\renewcommand{\thesection}{\arabic{section}.}
\renewcommand{\thesubsection}{\arabic{section}.\arabic{subsection}}
\renewcommand{\thesubsubsection}{\arabic{section}.\arabic{subsection}.\arabic{subsubsection}}
% make numbering of equation sectionwize
\catcode`\@=11
\@addtoreset{equation}{section}
\def\theequation{\arabic{section}.\arabic{equation}}
\catcode`@=12
\relax
\makeatletter
\def\section{\@startsection{section}{1}{\z@}{-3.5ex plus -1ex minus 
 -.2ex}{2.3ex plus .2ex}{\large}} 
\def\subsection{\@startsection{subsection}{2}{\z@}{-3.25ex plus -1ex minus 
 -.2ex}{1.5ex plus .2ex}{\normalsize\it}}
\def\subsubsection{\@startsection{subsubsection}{3}{\z@}{-3.25ex 
plus -1ex minus  -.2ex}{1.5ex plus .2ex}{\normalsize\it}}
\makeatother
\def\thefootnote{\fnsymbol{footnote}}

\begin{flushright}
hep-th/0405203\\
KUNS-1919\\
KEK-TH-959\\
May 2004
\end{flushright}
\vspace{1cm}
\begin{center}
\Large
Large $N$ limit of SYM theories with 16 supercharges\\
from superstrings on D$p$-brane backgrounds
\vspace{1cm}

\normalsize
Masako  {\sc Asano}
\footnote{
e-mail address:\ \ {\tt  asano@gauge.scphys.kyoto-u.ac.jp}}
 \quad and \quad 
 Yasuhiro {\sc Sekino} 
\footnote{
e-mail address:\ \ {\tt  sekino@post.kek.jp}}
\vspace{0.3cm}

${}^\ast$ 
%${}^1$
{\it Department of Physics, Kyoto University, \\
Kyoto 606-8502, Japan}

\vspace{0.4cm}

${}^\dagger$ 
%${}^2$
{\it Theory Division,
%Institute of Particle and Nuclear Studies\\
High Energy Accelerator Research Organization (KEK), \\
Tsukuba, Ibaraki 305-0801, Japan}

\vspace{0.7cm}
Abstract
\end{center}
We investigate the holographic correspondence between 
$(p+1)$-dimensional $(0\le p\le 4)$ SYM theories 
with 16 supercharges and superstring theories
on the near-horizon limit of D$p$-brane backgrounds.
Following an approach based on the tunneling picture,
we study Euclidean superstring semi-classically 
along null geodesics which connect
two points on the boundary of the spacetime.
We extend the analysis of hep-th/0308024 and
study the fermionic sector of the superstring.
For $p\ne 3$, we do not have world-sheet supersymmetry,
and the energies of  bosonic and fermionic fluctuations
do not match. 
By interpreting the superstring amplitudes as correlators 
of gauge theory operators with large R-charge $J$,  
we obtain gauge theory two-point functions 
including those of fermionic operators. 
Our approach yields results consistent with the 
previous supergravity analysis for the D0-branes,
including the subleading part in $J$. 
Our prediction from holography
is that the two-point functions for the supergravity 
modes are power-law behaved, even for the 
non-conformal ($p\ne 3$) SYM theories.

\newpage
\section{Introduction}
\def\thefootnote{\arabic{footnote}}
\setcounter{footnote}{0}

AdS/CFT correspondence \cite{Ma}, which states that $D$=4, ${\cal N}$=4
super Yang-Mills is equivalent to type IIB superstring on $\adss$
(near-horizon limit of D3-brane solution), is a remarkable
manifestation of the holographic principle~\cite{Su}. 
This provides a 
string representation of large $N$ gauge theory~\cite{Th}
in an explicit manner. 
We do not yet have complete understanding of the correspondence
since the superstring on $\adss$ has not been solved, but
substantial evidence 
and predictions for the strongly coupled
gauge theory have been given through the analysis of supergravity
theory (See \cite{AhGuMaOoOz, DhFr} for reviews).
Recently, there has been progress making use of the
semi-classical approximation of string theory.
Berenstein, Maldacena and Nastase (BMN) \cite{BeMaNa} considered
a point-like string rotating around $S^{5}$ with large angular momentum
($J$), and obtained the spectrum of quadratic fluctuations  
(or equivalently, considered superstring on the Penrose limit
of $\adss$). They identified the gauge theory operators
which correspond to the superstring states as  ones with large
R-charge, and predicted their anomalous dimensions.
The result of BMN have been checked by various gauge theory calculations
(See reviews \cite{SaSh, Pl}).
Moreover, strings in various stretched configurations
were studied, and the corresponding operators in the gauge theory
were identified (See~\cite{Ts}).

Holographic correspondence is expected to hold
for general D$p$-branes, not only for D3-branes. Namely,
it is conjectured that the  $(p+1)$-dimensional U($N$) SYM theories 
with 16 supercharges are dual to string theories on 
the near-horizon limit of the D$p$-brane solutions~\cite{ItMaSoYa}.
These SYM theories are not conformally invariant except for $p=3$.
We do not have powerful non-renormalization theorems resulting from the 
constraint of superconformal algebra, and little is known about
the strong coupling behavior of these theories.
It should be important to extend the well-established 
AdS/CFT correspondence to non-conformal cases and make 
definite predictions for these SYM theories.  

In~\cite{SeYo, Se}, holographic correspondence for the D0-branes
was studied in detail at the supergravity level.
Complete spectrum of the linearized supergravity on the 
near-horizon D0-brane background was worked out.  
Following the relation
between on-shell supergravity action and the generating functional
of the gauge theory correlators
proposed by Gubser, Klebanov and Polyakov~\cite{GuKlPo}
and by Witten~\cite{Wi} (GKP/W),
two-point functions of the (0+1)-dimensional SYM was obtained.
We can identify operators corresponding to supergravity modes
as the so-called Matrix theory currents \cite{TaRa}; 
state-operator mapping was determined with the aid
of the `generalized conformal symmetry' found by Jevicki
and Yoneya~\cite{JeYo}.
Prediction of \cite{SeYo, Se} is that
two-point functions corresponding to the supergravity modes are 
power-law behaved with fractional power in general.

In a previous paper~\cite{AsSeYo}, Yoneya and the present authors
started the investigation of the holography for D$p$-branes 
based on string theory.
We aimed at extending the results of BMN to general $p$,
taking an approach somewhat different from the one of BMN.
BMN read off the scaling dimensions of SYM operators
from the light-cone energy of string states. This is based on 
the following two assumptions: the spectrum on $\adss$ (where the energy is
defined with respect to the time in the global coordinate of AdS) 
is equal to the spectrum of SYM on $R\times S^{3}$ \cite{HoOo};
time translation operator of SYM on $R\times S^{3}$ 
generates the dilatation for SYM on a conformally equivalent
base space $R^{4}$.
It is not clear how to generalize the above statements
to $p\neq 3$, where we do not have 
the isometry of AdS space and the conformal invariance.

In~\cite{AsSeYo}, we have followed an alternative 
formulation of holography in a 
plane wave limit, proposed by 
Dobashi, Shimada and Yoneya for $p=3$ in~\cite{DoShYo}.
The idea is to consider Euclidean string theory  
in a spacetime where the `double Wick rotation' 
$(t\to -it, \psi\to -i\psi)$ is performed. 
Dominant trajectory which is called the `tunneling null 
geodesic' connects two points on the boundary \cite{DoShYo},
as opposed to the null geodesic in ordinary Minkowski $\adss$
which moves around the center of AdS. 
This allows us natural interpretation as holography.
String amplitude in the classical limit is essentially
equivalent to the geodesic approximation of the GKP/W prescription
in the large $J$ limit.
We study string around the tunneling null geodesic semi-classically.
Gauge theory correlators are given by the 
amplitudes of string theory.  Fluctuations around the tunneling null geodesic
are massive oscillators with the correct number of degrees of freedom; 
computation of the amplitude is a well-defined quantum mechanical problem.
Two-point functions obtained by this method reproduces 
the ones obtained by BMN \cite{AsSeYo}.

From the viewpoint of \cite{DoShYo}, it is obvious 
how to formulate holography for the case of general $p$ ($0\le p\le 4$).
Tunneling null geodesics exist in these D$p$-brane backgrounds
after the double Wick rotation, and we can 
perform the semi-classical analysis of the string amplitude.  
A difference from the $p=3$ case is that string in the
semi-classical limit has (world-sheet) time-dependent masses.
In \cite{AsSeYo}, we have developed a general method for
the quantization of time-dependent harmonic oscillators, 
and studied the amplitude for the bosonic sector. 
We have obtained the two-point functions of SYM operators 
corresponding to the lowest modes (supergravity modes), and confirmed
that they are consistent in the large $J$ limit 
with the results of \cite{SeYo} based on the GKP/W prescription. 
We have discussed the
qualitative behavior of the correlators corresponding to the
string higher excitations. These correlators show non-trivial
behavior: For $p<3$, we have massive-like exponential correlation
functions.

The main purpose of the present paper is to study the fermionic sector of the 
superstring, which was left for future work in \cite{AsSeYo}.
Contrary to the 
$p=3$ case where the superstring in the
semi-classical limit has world-sheet supersymmetry~\cite{Me,
CvLuPoSt}, the superstring for $p\ne 3$ has no such supersymmetry;
mass of the fermions and bosons do not match.
We obtain the amplitude for the supergravity modes, and
discuss consistency with the correlation functions
of fermionic operators obtained from the supergravity analysis in 
\cite{Se}. We also determine the contribution to the amplitude 
from the zero-point energies of 
bosonic and fermionic fluctuations; by including this, 
the correlators obtained from string theory agree
with the ones obtained from 
the supergravity analysis, to the sub-leading order in $J$.
We also comment on the quantization of string higher modes.
The analysis of this paper is performed in a gauge condition
different from the conformal gauge used in \cite{AsSeYo}.
In this gauge, superparticle action on the near-horizon D$p$-brane 
background becomes formally equivalent to the one on
${\rm AdS}_{p+2}\times S^{8-p}$,
and the masses of supergravity modes become constant.
We will give a brief summary of the analysis of~\cite{AsSeYo}
in the present gauge condition.

The organization of this paper is as follows. 
Section 2 is a review of our formalism and the
results of the previous paper~\cite{AsSeYo}.
We introduce the double Wick rotation of the near-horizon
D$p$-brane backgrounds, and see that 
null geodesics connect two points on the boundary. We then perform 
semi-classical expansion, and obtain
amplitude for the bosonic fluctuations.
After determining the state-operator correspondence,
we predict the two-point functions.  
In section 3, we derive the action for the fermionic
fluctuation in the semi-classical limit.
In section 4, we study superstrings and obtain 
two-point functions for the
fermionic sector. Amplitudes and two-point functions
for the supergravity mode are computed in 4.1. 
Consistency with the previous
supergravity analyses of~\cite{SeYo, Se} is examined in 4.2. 
Quantization of string higher modes is briefly discussed in 4.3.
We conclude in section 5 and mention possible directions 
for further studies. We collect a few technical details 
in appendices.
In Appendix A, we derive the superstring action in the 
conformal gauge, and explain its relation 
with the action used in the text.  
In Appendix B, we describe the
relation between the Poincar\'{e} and the global
coordinates; the latter coordinate system is employed  
for the semi-classical expansion in section 2 and 3. 
In Appendix C, we discuss the supersymmetry and the 
$\kappa$-symmetry in the semi-classical limit. 

%%%%%%%%%%%%%%%%%%%%%%%%%%%%%%
\section{Tunneling null geodesics and holography for D$p$-branes}
%\label{sec2} 
%%%%%%%%%%%%%%%%%%%%%%%%%%%%%%
%%%
\subsection{D$p$-brane solution in the near-horizon limit}
%%%

D$p$-brane solution of type II supergravity 
in the near-horizon limit 
is characterized by the following metric, 
Ramond-Ramond $(p\!+\!2)$-form field strength and the dilaton:
\begin{eqnarray}
ds^{2}&=&L^{2} \left\{H^{-1/2}(-dt^{2}+d x_{a}^{2})+H^{1/2}
(dr^{2}+r^{2}(d\theta^2+ \cos^2\theta d\psi^2 + \sin^2\theta  
d\Omega_{6-p}^{2}))\right\},
\nonumber\\
F_{p+2}& =& L^{p+1}\partial_{r} H^{-1}dt\wedge dx_{1}\wedge \cdots
\wedge 
dx_{p}\wedge dr,
\label{eq:nhDp}
\\
%e^{\phi}&=& g_{s}e^{\tilde\phi},\quad e^{\tilde\phi}=H^{\frac{3-p}{4}},
&& \quad  e^{\phi}= g_{s}H^{\frac{3-p}{4}},
\quad H = {1\over r^{7-p}} 
\nonumber
\end{eqnarray}
where $a=1,\ldots, p$, $L=q_{p}^{1/(7-p)}$ and 
$q_{p}=\tilde{c}_{p}g_{s}N\ell_{s}^{7-p}$ with 
$\tilde{c}_{p}=2^{6-p}\pi^{(5-p)/2}\Gamma{(7-p)/2}$.
Integer $N$ is the number of the D$p$-branes.
The coordinates in (\ref{eq:nhDp}) are dimensionless;
we have rescaled the
usual coordinates with length dimension 1 as
$(t,x_{a},r)\to L(t,x_{a},r)$. 
We will use this convention throughout this paper%
\footnote{This is slightly different
from the convention of the previous paper \cite{AsSeYo},
eq.(2.1).}.
Note that this background is conformally equivalent to 
AdS$_{p+2}\times S^{8-p}$ 
in the sense that the metric is written as 
$ds^2 =L^{2}r^{(p-3)/2}\, ds^2_{{\rm AdS}_{p+2}\times S^{8-p}}$.

Superstring or supergravity theory on the background (\ref{eq:nhDp})
is expected to be dual to the $(p+1)$-dimensional U($N$) SYM 
with 16 supercharges.
According to Gubser, Klebanov and Polyakov~\cite{GuKlPo} and 
Witten~\cite{Wi} (GKP/W), correlation functions of 
certain operators of $D$=4, ${\cal N}$=4 SYM 
are given by perturbative analysis of supergravity
on $\adss$. 
We assume that the GKP/W's relation between 
supergravity action and the generating functional of the correlators:
\begin{equation}
 e^{-S_{SG}[\phi_{0}]}=\langle e^{\int d^{p+1}x \phi_{0}(x) 
{\cal O}(x)}\rangle
\label{eq:GKP}
\end{equation} 
holds for general D$p$-branes. 
Here, $S_{SG}$ is a classical value of the supergravity action 
on the near-horizon D$p$-brane background; supergravity fields
are on-shell with a boundary condition  $\phi=\phi_{0}$ imposed
at the $(p+1)$-dimensional `boundary' of the ${\rm AdS}_{p+2}$-like 
space. More precisely, we put the boundary near the end of the 
near-horizon region $r=r_{\Lambda}\sim 1$, as in \cite{GuKlPo}. 
Also, we keep in mind that the prescription (\ref{eq:GKP}) 
is formulated in the Euclidean spacetime where the time
of the background and the gauge theory are Wick rotated. 
The operators ${\cal O}$ in $D$=4, ${\cal N}$=4 SYM which correspond
to the supergravity modes are the BPS operators \cite{AhGuMaOoOz, DhFr}.
The operators ${\cal O}$ for the SYM theories in general dimensions
should be their analogs: for $(0+1)$-dimensional SYM, we assume
${\cal O}$ are the Matrix theory currents obtained by Taylor and
Van Raamsdonk from the
one-loop analysis of BFSS Matrix theory~\cite{TaRa}; for other
dimensions they will be the ones given in~\cite{TaRa2} which are
obtained by T-dualities from the Matrix theory currents. 

Perturbative string or supergravity theory should
be a good description when string coupling and the background
curvature are small. For $p=3$, conditions for the weak coupling 
and small curvature are $g_{s}\ll 1$ and $g_{s}N\gg 1$,
that is, they are satisfied in
the large $N$ limit with large 't Hooft coupling. 
For $p\neq 3$, the curvature 
and the string coupling depend on the radial coordinate $r$. 
Curvature radius $R_{c}$ for the background is given as
$R_{c}\sim LrH^{1/4}\sim Lr^{(p-3)/4}$; curvature is 
small in string unit $(R_{c}\gg \ell_{s})$ when
%\begin{eqnarray}
%r&\ll& (g_{s}N)^{4\over (7-p)(3-p)}\quad (p< 3),\nonumber\\
%r&\gg& (g_{s}N)^{4\over (7-p)(3-p)}\quad (p> 3).
%\label{eq:smallcurv}
%\end{eqnarray}
\begin{equation}
r\ll (g_{s}N)^{4\over (7-p)(3-p)}\quad (p< 3),\qquad
r\gg (g_{s}N)^{4\over (7-p)(3-p)}\quad (p> 3).
\label{eq:smallcurv} 
\end{equation}
String coupling is small 
%$(e^{\phi}=g_{s}r^{-(3-p)(7-p)/4}\ll 1)$ 
$(e^{\phi}\ll 1)$ when
\begin{equation}
g_{s}^{(3-p)(7-p)\over 4}\ll r \quad (p< 3),\qquad
g_{s}^{(3-p)(7-p)\over 4}\gg r \quad (p> 3).
\label{eq:weakcoup}
\end{equation}
%\begin{eqnarray}
%&&g_{s}^{(3-p)(7-p)\over 4}\ll r \quad (p< 3),\nonumber\\
%&&g_{s}^{(3-p)(7-p)\over 4}\gg r \quad (p> 3).
%\label{eq:weakcoup}
%\end{eqnarray}
If we have $g_{s}N\gg 1$ and $g_{s}\ll 1$,
or $N\to \infty$ with $g_{s}N$ fixed and large,
the region where (\ref{eq:smallcurv}) and (\ref{eq:weakcoup})
are valid becomes large enough to cover the whole near-horizon
region  $(0\le r\le 1)$ except the origin $r=0$.  
Thus we will be able to explore the gauge theory in this regime 
(large $N$ limit with strong effective coupling) 
by supergravity or superstring theory on this background
via holographic correspondence%
\footnote{If we approach arbitrarily
close to the origin $r=0$, string coupling (for $p<3$) or
curvature (for $p>3$) becomes strong. However, since
we are using essentially the information 
far from the origin in the bulk computation, 
we believe that this condition is sufficient.}. 
Supergravity analysis of the holographic correspondence for
D0-branes has been previously performed. 
In \cite{SeYo}, two-point functions of (0+1)-dimensional SYM
for bosonic operators were obtained by using (\ref{eq:GKP}). 
In \cite{Se}, those for fermionic operators were obtained
following the fermionic extension of the GKP/W prescription
\cite{HeSf}.

We would like to recast the GKP/W prescription in the `first-quantized'
picture. 
We take the following approach first proposed in \cite{DoShYo} 
and developed in \cite{AsSeYo}: We perform `double Wick rotation' 
for time and angle 
\begin{equation}
t\to -it,\quad \psi\to -i\psi  
\end{equation}
in the background (\ref{eq:nhDp}), and study the amplitude 
of superstring with Euclidean world-sheet time $\tau$.
The Wick rotation of $t$ is natural, recalling the fact that
the GKP/W prescription was formulated in the Euclidean signature.
As we see shortly, after the double Wick rotation, 
we have null geodesics
(which we call `tunneling null geodesics')  connecting two points 
on the boundary% 
\footnote{
Null geodesics in the D$p$-brane background without double Wick rotation
never reach the boundary for $0\le p\le 4$. 
Thus, the method of BMN, which is based on an 
expansion around Minkowski null geodesics, does not seem 
to have direct relation with GKP/W prescription. 
In addition, for $p\ne 3$, null geodesic falls into 
a singular point $r=0$.}
for $0 \le p \le 4$. Fluctuations around it have the same
number of degrees of freedom as the theory without double
Wick rotation.
We will obtain the amplitude semi-classically,
and interpret it as correlators in the gauge theory.

\subsection{Classical limit: tunneling null geodesic}

Let us consider the bosonic part of the superstring action
\begin{equation}
 S_{b}={1\over 4\pi}\int d\tau \int_{0}^{2\pi\tilde\alpha} d\sigma
\sqrt{h} h^{\alpha\beta}\partial_{\alpha}x^{\mu}
\partial_{\beta}x^{\nu} \tilde{g}_{\mu\nu}
\label{eq:Sb}
\end{equation}
where $\alpha,\beta$ denote the world-sheet coordinates
$\tau,\sigma$ with signature $(+,+)$. We shall set $\alpha'=1$
hereafter.
The metric $\tilde{g}_{\mu\nu}$ is given by (\ref{eq:nhDp})
after double Wick rotation:
\begin{eqnarray}
d\tilde{s}^2 &=& 
L^2 r^{-{3-p \over 2}}
\Big[
\left(\ft{2}{5-p}\right)^{2}\left( dt^{2}+dx_{a}^{2}+dz^{2}\over z^{2}\right)
+\big( d\theta^2 - \cos^2\theta d\psi^2 + \sin^2\theta  
(d\Omega^{(6-p)}_l)^{2}\big)
\Big]\nn\\
\label{eq:poincare}
\end{eqnarray}
where the radial coordinate $z$ of the Poincar\'{e} coordinate
is defined by
\begin{equation}
z={2\over 5-p}r^{-(5-p)/2}.
\label{eq:relationzr}
\end{equation}
In the following analysis, we represent the AdS$_{p+2}$ in 
the global coordinates%
\footnote{We are using the word `global' in a formal sense.
D$p$-brane geometry for $p\ne 3$
(before double Wick rotation) cannot be extended past 
the singularity $r=0$. 
Note that we are always considering a single
Poincar\'{e} patch in our approach.} $(\rho, s, \Omega_{p})$:
\begin{eqnarray}
d\tilde{s}^{2}&=&L^2 r^{-{3-p \over 2}}
\Big[
\left(\ft{2}{5-p}\right)^{2}\left(
 d\rho^2+\sinh^2\!\rho \, (d\Omega_i^{(p)})^2 +\cosh^2\!\rho\, ds^2\right)
\nonumber\\
&& \hspace*{4cm}+
\big( d\theta^2 - \cos^2\theta d\psi^2 + \sin^2\theta  
(d\Omega^{(6-p)}_l)^{2}\big)
\Big]
\label{eq:metricDpdW}
\end{eqnarray}
Relation between two coordinate systems is described in Appendix B.
Here we note that $z$ is written as
\begin{equation}
z =
\frac{\tilde{\ell} }
{\cosh\rho \, \cosh\!s -\sinh\rho \, \Omega^{(p)}_{p+1}}
\end{equation}
where $\tilde{\ell}$
% ={2\over 5-p} \ell$ ($ \ell\! =\! J/E$) 
is a parameter which specifies the relation between 
Poincar\'{e} and global coordinates. It is introduced 
in order to represent a one-parameter family of 
geodesics in the Poincar\'{e} coordinates by a
single one, (\ref{eq:trajG}) below, in the global coordinates.

Let us consider a point-like ($\sigma$-independent) classical 
solution with $\rho= \theta=0$. It is convenient to fix the
parameterization of the world-sheet such that $\sqrt{h}h^{\tau\tau}$
`absorbs' the Weyl factor of the background metric%
\footnote{Somewhat similar gauge condition is used for the
analysis of string in AdS space in~\cite{MeThTs, DaMaWa}.
There, $\sqrt{h}h^{\tau\tau}$ was chosen to 
absorb the $1/z^{2}$ factor of the AdS metric in the
Poincar\'{e} coordinates.}:
\begin{equation}
\sqrt{h} h^{\tau\tau} =\bar{r}(\tau)^{3-p \over 2},
\end{equation} 
where $\bar{r}(\tau)$ is a function of $\tau$ defined by
\begin{equation} 
\bar{r}(\tau)=
\left(\frac{\cosh\!\tau}{\ell} \right)^{2/(5-p)},
\quad
\ell= {5-p\over 2}\,\tilde{\ell}\,.
\label{eq:defbarr}
\end{equation} 
Classical solution, which satisfies the massless condition, is given by
\begin{equation}
\psi = \ft{2}{5-p} \tau \, ,\quad s=\tau. 
\label{eq:trajG}
\end{equation}
We have a conserved angular momentum along $\psi $: 
\begin{equation}
J\equiv L^2 \dot{\psi} \sqrt{h} h^{\tau\tau} r^{-{3-p \over 2}}
{\tilde\alpha}
\label{eq:defJ}
\end{equation}
where dot $(\,\dot{}\,)$ means the derivative with respect to $\tau$.
(We will also use $'$ for $\partial_{\sigma}$.)
We set the world-sheet length $\tilde\alpha$ as
\[
 \tilde\alpha={5-p\over 2}{J\over L^{2}},
\]
so that the solution (\ref{eq:trajG}) has angular momentum $J$.
%Here, we assume that $J$ takes integer values, since $\psi$ was originally
%an angular variable with period $2\pi$. 
%%%%%%%%%%%%%%%%%%%%%%%%%
\begin{figure}[t]
   %\hspace*{4cm}
   \begin{center}
     \vspace*{-5mm}
   \epsfxsize=7.5cm
   \epsfbox{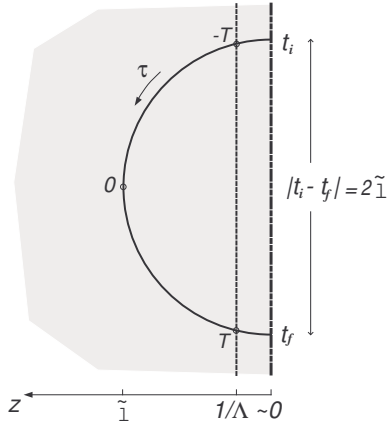}
     \vspace*{-8mm}
   \end{center}
\caption[x]{\small
The null geodesic 
$t^2+z^2=\tilde{\ell}^2$
(\ref{eq:trajP}) 
in the double Wick rotated D$p$-brane ($0\le p\le 4$) background 
(\ref{eq:metricDpdW}).
It connects two points on the $(p+1)$-dimensional boundary
(at $z\sim 0$) of the ${\rm AdS}_{p+2}$-like space.
Cutoff is introduced at a radial position $z=1/\Lambda$.} 
\end{figure}

In the Poincar\'{e} coordinates, the above solution takes the form
\begin{equation}
 z={\tilde\ell\over \cosh \tau},\quad 
t=\tilde\ell \tanh \tau,\quad \psi={2\over 5-p}\tau,
\label{eq:trajP}
\end{equation}
as we explain in Appendix B. We see that this geodesic 
connects two points 
on the ($p+1$)-dimensional boundary at $z\sim 0$ (See Figure 1.); 
it starts from $t=t_{i}$ at $\tau=-\infty$ and reaches
$t=t_{f}$ at $\tau=\infty$. The 
separation between the endpoints is
\begin{equation}
 |t_{f}-t_{i}|=2\tilde{\ell}.
\label{eq:tfti}
\end{equation} 
For future purpose, we set a cutoff at 
$z(\tau =\pm T)=1/\Lambda$. We will take the
cutoff large in the sense that $\tilde\ell\gg 1/\Lambda$,
so that large portion of the geodesic lies in the region% 
%\footnote{We should keep in mind that we cannot take 
%$\Lambda$ arbitrary large since the near-horizon 
%approximation to the D$p$-brane 
%geometry is valid when $1\ge z$. 
%However, in the limit $\tilde\ell\Lambda\to \infty$, 
%we can think of $\Lambda$ as infinity in most computations.
\footnote{We should keep in mind that we cannot take 
$\Lambda$ arbitrarily large since the near-horizon 
approximation of the geometry is valid when $z\ge 1$. 
However, we also note that 
the coordinate separation in gauge theory $|t_{i}-t_{f}|$ 
is much larger than the cutoff scale $1/\Lambda$
when $\tilde\ell\Lambda\to \infty$.} $z\ge 1/\Lambda$. Note that 
the proper-time cutoff $T$ and the radial cutoff $\Lambda$
are related by  
\begin{equation}
2\tilde\ell\Lambda\sim e^{T}.
\label{eq:lambdat}
\end{equation}
when $T\to \infty$ (or $\tilde\ell\Lambda \to \infty$).

Amplitude in the classical limit is given by the
classical value of the action as $e^{-S^{(0)}}$. 
In our case, $S^{(0)}$ is zero, but
we assume here that the initial 
and final states are in the
momentum representation with respect to $\psi$. 
We have a factor 
$e^{-J\psi(T)+J\psi(-T)}=e^{-J\int_{-T}^{T}d\tau \dot{\psi}}$
from the Fourier transformation (in the Euclidean formulation).
Classical limit of such an amplitude
is given by $e^{-\bar{S}}$ where $\bar{S}$ is the Routh function%
\footnote{Routh function was used  to obtain the classical 
limit of the amplitude with fixed momentum,
in the context of the D-particle scattering \cite{BeBePoTs}.}: 
\begin{equation}
 \bar{S}=S-\int_{-T}^{T} d\tau J\dot\psi={4\over 5-p}JT.  
\label{eq:barS}
\end{equation} 
We can also say that $\bar{S}$ is the length of the geodesic
(spacelike geodesic) in the Euclidean 
$(p+2)$-dimensional subspace:
\begin{equation}
 \bar{S}=L^{2}\int^T_{-T}\! d\tau
\int_0^{2\pi \tilde{\alpha}}\!\!\! d\sigma 
\sqrt{h}h^{\tau\tau} \left(\ft{2}{5-p}\right)^2 \dot{s} r^{-{3-p\over 2}}. 
\end{equation} 
%We will interpret the amplitude including the contribution 
%as a gauge theory correlator.
The amplitude  $e^{-\bar{S}}$ is the one obtained through
geodesic approximation of the GKP/W prescription~\cite{SuWi}.

Note that $J$ takes integer value since this is the
conjugate momentum for $\psi$ which was originally a
periodic variable with period $2\pi$. 
Also note that the fermionic fluctuations, whose 
spectrum will be studied later, have spin angular momentum.
When we consider excitations of fermionic oscillator,
we must take into account the spin contribution $1/2$ for one
excitation in the Fourier transformation factor. 
That is, when $N_{f}$ fermionic 
oscillators are excited, we have to use
\begin{equation}
 \bar{S}\to {4\over 5-p}(J+{1\over 2}N_{f})T
\end{equation}
instead of $\bar{S}$ in (\ref{eq:barS}).
We will interpret the product of the above contribution
and the amplitudes for  the fluctuations as two-point functions
of the gauge theory.

%%%
\subsection{Bosonic fluctuations around the tunneling null geodesic}
\label{sec:2.3}  
%%%

We study the fluctuations around the tunneling null geodesic
in the semi-classical expansion, which is valid in the 
large $L$ (or $J$) limit. We expand the fields
around the classical trajectory (\ref{eq:trajG}) 
in powers of $1/L$:
\begin{eqnarray}
\psi &=& {2\over 5-p}\,\tau +{\psi^{(1)} \over L} +{\psi^{(2)} \over L^2}+\cdots,
\quad
s \,=\, \tau +{s^{(1)} \over L} +{s^{(2)} \over L^2}+\cdots,
\nonumber\\
\rho &=& {\rho^{(1)} \over L} +\cdots, \quad 
\theta \,=\, {\theta^{(1)} \over L} +\cdots.
\label{eq:scexpand}
\end{eqnarray}
We substitute this into the action (\ref{eq:Sb}) with the
gauge condition
\begin{equation}
 h^{\tau\sigma}=0, \quad \sqrt{h}h^{\tau\tau}
=\bar{r}(\tau)^{3-p\over 2} \left(
={1\over \sqrt{h}h^{\sigma\sigma}} \right).
\label{eq:hgauge}
\end{equation} 
The bosonic action is written as 
\begin{equation}
 S_b= L^2\left(
S_{b}^{(0)} + {1\over L} S_{b}^{(1)} + {1\over L^2} S_{b}^{(2)} +\cdots
\right).
\end{equation}
The first two terms $S_b^{(0)}$ and $S_b^{(1)}$ vanish due to 
the classical equation of motion.

By using the Virasoro constraint 
\begin{equation}
\partial_{\tau}x^{\mu}\partial_{\tau}x^{\nu}\tilde{g}_{\mu\nu} 
- {h^{\sigma\sigma} \over h^{\tau\tau}}
\partial_{\sigma}x^{\mu}
\partial_{\sigma}x^{\nu}\tilde{g}_{\mu\nu}=0,\quad
\partial_{\tau}x^{\mu}
\partial_{\sigma}x^{\nu}\tilde{g}_{\mu\nu}=0
\label{eq:virasoro}
\end{equation}
at order ${\cal O}(L^{-1})$ 
\begin{equation}
\ft{2}{5-p} \partial_\alpha s^{(1)} -\partial_{\alpha}\psi^{(1)}=0 \quad (\alpha=\tau, \sigma),  
\end{equation}
the action $S_b^{(2)}$ for the fluctuations 
at the leading order becomes%
\footnote{In fact, 
the action (\ref{eq:actionbos2}) can be formally obtained
by first writing the action for the fluctuations
around null geodesic in the background
without the double Wick rotation~\cite{GiPaSo, FuItSe}, 
and then replacing $\tau\to -i\tau$ in that action.
% (which contain only the physical degrees of freedom). 
This is true also for the fermionic action obtained in Section 3.}
\begin{eqnarray}
S_b^{(2)}={1\over 4 \pi}
\int d\tau \int_0^{2\pi \tilde{\alpha}  } \!\!\! d\sigma\,
\Big[
\left(\ft{2}{5-p}\right)^2 (\dot{x}_i^2+\bar{r}(\tau)^{p-3} x_l'^{\,2})
+\dot{y}_l^2+\bar{r}(\tau)^{p-3} y_l'^{\,2}
\nonumber\\
\hspace*{5cm}
+\left(\ft{2}{5-p}\right)^2
(x_i^2 +y_l^2)
\Big] \quad +\, {\cal O}(L^{-1}).
\label{eq:actionbos2}
\end{eqnarray}
Here we have defined 
$x_{i}$ $(i=1,\ldots,p+1)$ and $y_{l}$ $(l=1,\ldots 7-p)$
by
\begin{equation}
 x_{i}=\rho^{(1)} \Omega_{i}^{(p)},\quad 
 y_{l}=\theta^{(1)} \Omega_{l}^{(6-p)}.
\label{eq:xycoord}
\end{equation}
The equations of motion for $x_i$ and $y_l$ are given respectively as
\begin{eqnarray}
&& \ddot{x}_i + \bar{r}(\tau)^{p-3} x_i{''} - x_i = 0 ,
\nonumber\\
&& \ddot{y}_l + \bar{r}(\tau)^{p-3} y_l{''} - 
\left( \ft{2}{5-p} \right)^2
y_l = 0.
\label{eq:xyeom}
\end{eqnarray}
As explained in Appendix~A, the action (\ref{eq:actionbos2})
is related by field redefinitions and rescaling of $\tau$
to the action (\ref{eq:actionbosconf}) derived in the conformal gauge.

\subsection{Amplitude and correlation functions: bosonic supergravity modes}
The equations of motion (\ref{eq:xyeom}) for the supergravity modes
(lowest excitations, $\partial_\sigma=0$)
are those of harmonic oscillators with constant mass:
\begin{equation}
 \ddot{X}-m^{2}X=0,
\end{equation}
where $m=m_{x(p)}\equiv 1$ for $x_{i}$, and
$m=m_{y(p)}\equiv {2\over 5-p}$ for $y_{l}$. 
The Hamiltonian for such an oscillator is
\begin{equation}
H={m \over 2} \left( a^{\dagger}a+ {1\over 2} \right), 
\end{equation}
and the S-matrix from $\tau=-T$ to $\tau=T$ is given by%
\footnote{
If we have started from the string action in the conformal gauge,
the supergravity modes would have 
time-dependent mass. As shown in the previous paper \cite{AsSeYo}, 
we can quantize
the time-dependent oscillators and obtain the 
result equivalent to (\ref{eq:Sbosconstm}).}
\begin{equation}
\exp \left( -\int^T_{-T} d\tau H \right) 
= e^{-m (2a^{\dagger}a+1)T }. 
\label{eq:Sbosconstm}
\end{equation}

Combining the classical result
$\exp (-\bar{S})$ and the factors from the 
oscillators (\ref{eq:Sbosconstm}), the S-matrix when
supergravity states are excited becomes
\begin{eqnarray}
  S(T) &=& e^{-\bar{S}}e^{-2\left(m_{x(p)}N^{x}_{0}
+m_{y(p)}N^{y}_{0}+c_{0}\right)T}
\label{eq:sugraamp}
\end{eqnarray}
where $N_{0}^{x}$ and $N_{0}^{y}$ are occupation numbers
of the $x_{i}$ and $y_{l}$ oscillators for the supergravity modes.
Constant $c_{0}$ is the sum of the zero-point energies;
this will be determined later in section 4, taking 
into account the contribution from the fermionic fluctuations.

Using (\ref{eq:tfti}) and (\ref{eq:lambdat}),
the S-matrix (\ref{eq:sugraamp}) can be expressed in the form
\begin{equation}
 S(T)\sim \left( |t_f-t_i|\Lambda \right)^{-{4\over 5-p}J-
2N^{x}_{0}-{4\over 5-p}N^{y}_{0}-2c_{0}},
\label{eq:twopointb}  
\end{equation}
We interpret this S-matrix  
as a two-point function in the gauge theory  
$\langle{\cal O}(t_f){\cal O}(t_i)\rangle$. 
An operator ${\cal O}$ which corresponds to 
each state constructed by the above oscillators 
can be identified, by extending naturally 
the proposal of BMN \cite{BeMaNa}:
Ground state (classical limit) corresponds to ${\rm Tr} Z^{J}$,
where $Z$ is given by the linear combination $Z=\phi_{8}+i\phi_{9}$
of two of the $(9-p)$ scalars of $(p+1)$-dimensional SYM.
Excitations of $y_l$ oscillators correspond to the insertions of
the remaining $(7-p)$ scalars $\phi_{l}$ into the trace; 
excitations of $x_i$ oscillators correspond to
the insertions of the covariant derivatives $D_{i}$. 
For example, excitations of a single bosonic supergravity oscillator 
$a^{\dagger}_{0,i}|0\rangle$ 
and $a^{\dagger}_{0,l}|0\rangle$ correspond to 
\begin{equation}
{\cal O}_i \sim {\rm STr} (Z^{J-1} D_{i}Z +\cdots)
\quad
\mbox{and}
\quad
{\cal O}_l \sim {\rm STr} (Z^{J} \phi_l +\cdots),
\end{equation}
respectively.
%\begin{equation}
%\langle{\cal O}(x_f){\cal O}(x_i)\rangle
%\sim 
% \left( |x_f-x_i|\Lambda \right)^{-\left({4 \over 5-p} J +m (2N+1) \right)} 
%\end{equation}
%where $m=1 (\equiv m^x_{b(p)}) $ for ${\cal O}_i$ 
%and $m=2/(5\!-\!p)(\equiv m^y_{b(p)}) $ for ${\cal O}_l$.
%We have identified the operators following the discussion given 
%by \cite{BeMaNa} in the case of  $p=3$.
%The result of course agrees with that given in \cite{AsSeYo}
%where the action in conformal gauge is analyzed. 

%%%%%%%%%%%%%%%%%%%%%%%%%%%%%%%%%%%%
%%%%%%%%%%%%%%%%%%%%%%%%%%%%%%%%%%%%
\section{Action of the fermionic fluctuations in the semi-classical
limit}
%%%%%%%%

Having reviewed our formalism for obtaining the gauge theory
correlators from the string amplitudes, 
we shall now study the fermionic sector. This section is devoted to
the derivation of  the action for the fermionic 
fluctuations at the leading order in $L\to \infty$.
We start from the superstring action in the general backgrounds
to the quadratic order in fermions,  
following the convention of 
Cvetic et al. \cite{CvLuPoSt}.

%%%
\subsection{type IIA}
%%%
Type IIA Green-Schwarz action at the quadratic order in fermions
in the Minkowski signature is 
\begin{equation}
S^{(M)}_{f,A} 
= -{i\over 2\pi} \int d\tau \int^{2\pi\tilde{\alpha}}_{0} \! d\sigma \, 
\Theta^{T}\Gamma_0
(\sqrt{-h}h^{\alpha\beta}-\epsilon^{\alpha\beta}\Gamma_{11} )
\partial_{\alpha}x^{\mu}\Gamma_\mu  {\cal D}_{\beta} \Theta .
\end{equation}
Here 
%The derivative ${\cal D}_{\alpha}\Theta$ is defined by
\begin{equation}
 {\cal D}_{\alpha}\Theta=\nabla_{\alpha}\Theta 
+\partial_{\alpha}X^{\mu} \Omega \Gamma_{\mu}\Theta,
\end{equation}
\begin{equation}
\Omega=- {1\over 16} e^\phi\, \Big(\Gamma_{11}\,  
\Gamma^{\rho\sigma}\, 
F_{\rho\sigma} - {1\over 12} \Gamma^{\rho\sigma\lambda\tau}\, 
F_{\rho\sigma\lambda\tau}\Big)
\label{eq:Omega}
\end{equation}
for the D$p$-brane background which has RR two-form (or four-form)
field strength $F_{\rho\sigma}$ (or $F_{\rho\sigma\lambda\tau}$). 
The fermionic fields $\Theta$ are 32-component Majorana spinors in
the real representation $\Theta^{*}=\Theta$, and 
$\nabla_{\alpha}\Theta=\partial_{\alpha}\Theta 
+{1\over 4}\partial_{\alpha}X^{\mu}
\omega_{\mu}{}^{\hat{\nu}\hat{\rho}}\Gamma_{\hat{\nu}\hat{\rho}}
\Theta $ is the covariant derivative.

As we have done for the bosonic part, we substitute the 
near-horizon D$p$-brane background into the action, 
perform double Wick rotation, 
%background
%(\ref{eq:metricDpdW}) 
and expand the action around the classical trajectory (\ref{eq:trajG}).
% as in (\ref{eq:scexpand}) 
%to obtain the action 
%for fluctuations in the limit $L\to \infty$.
We take the vielbein as follows:
%To avoid confusion about the representation of spinors or Gamma matrices,
%we assume that the Wick rotation is only performed for 
%space-time indices and not for indices of local Lorentz frame;
%We take the vielbein frame
\begin{eqnarray}
e^{\hat{s}} &\!=\!& -i \ft{2}{5-p} L r^{-{3-p \over 4}} 
\cosh\!\rho \,ds ,\quad 
\quad 
e^{\hat{\psi}} = -i L r^{-{3-p \over 4}} \cos\!\theta \, d\psi, 
\nonumber\\
e^{\hat{\rho}}&\! =\! & \ft{2}{5-p} L r^{-{3-p \over 4}} d\rho ,\quad
e^{\hat{\alpha}_a} =    \ft{2}{5-p}  L r^{-{3-p \over 4}}
 \sinh\!\rho 
 \left(\prod_{k=1}^{a-1} \sin\!\alpha_k \!\right)\!
d\alpha_a,  
\nonumber\\
e^{\hat{\theta}} &\!=\!& L r^{-{3-p \over 4}} d\theta,\quad
e^{\hat{\chi}_m} = L r^{-{3-p \over 4}}
 \sin\!\theta
 \left(\prod_{k=1}^{m-1} \sin\!\chi_k \! \right)\!
d\chi_m 
 \label{eq:vielbein} 
\end{eqnarray}
and represent  the double Wick-rotated metric (\ref{eq:metricDpdW}) as
\begin{equation}
d\tilde{s}^{2} =  - e^{\hat{s}} e^{\hat{s}} + e^{\hat{\psi}}e^{\hat{\psi}}
+ e^{\hat{\rho}}e^{\hat{\rho}} + e^{\hat{\theta}}e^{\hat{\theta}}
+\sum_{a=1}^p e^{\hat{\alpha}_a}e^{\hat{\alpha}_a}  
+ \sum_{m=1}^{6-p} e^{\hat{\chi}_m}e^{\hat{\chi}_m} .
\label{eq:metricviel} 
\end{equation}
where $\{\alpha_a\}$ and  $\{\chi_{m}\}$ are angular coordinates
of $S^{p}$ and $S^{6-p}$, respectively.
Note that we have the $-i$ factors in the first line 
of (\ref{eq:vielbein}) because of the double Wick rotation.
Local Lorentz frame has the usual Minkowski signature:
$e^{\hat{s}}=- e_{\hat{s}}$,
$e^{\hat{\psi}} = e_{\hat{\psi}}$.
The components of the
gamma matrices $\Gamma_{\hat{\mu}}$ are real.

Euclidean action obtained by the world-sheet Wick rotation
$\tau\to -i\tau$ is
\begin{equation}
S_{f,A} 
= {i\over 2\pi} \int d\tau \int^{2\pi \tilde{\alpha}}_{0}
d\sigma \Theta^{T}\Gamma_0
(\sqrt{h}h^{\alpha\beta}-i\epsilon^{\alpha\beta}\Gamma_{11} )
\partial_{\alpha}X^{\mu}{\Gamma}_\mu {{\cal D}}_{\beta} \Theta. 
\end{equation}
We fix the world-sheet metric as in the bosonic case (\ref{eq:hgauge}).
To obtain the action at the leading order in $L\to \infty$, 
we note
\begin{equation}
 \partial_{\alpha} X^{\mu} \Gamma_{\mu} =
 \partial_{\alpha} X^{\mu} e^{\hat{\nu}}_{\mu}\Gamma_{\hat\nu} =
-i \ft{2}{5-p} L \bar{r}^{-{3-p \over 4}} 
(\Gamma_{\hat{\psi}} + \Gamma_{\hat{s}} )\delta_{\alpha\tau} \quad
+\,{\cal O}(L^0).
\end{equation}
Also, we have  the RR field strength, given by (\ref{eq:RRDpg})
with the replacement $s\to -is$: 
\begin{eqnarray}
\tilde{F}_{p+2} &=& i L^{p+1}(7-p) \left(\ft{2}{5-p}\right)^{p+2}
r^{(3-p)^2 \over 2} \sinh^p\!\rho \cosh\!\rho \,ds\wedge dv^{p} \wedge d\rho
\label{eq:RRdW}
\\
&=& i \, (7\!-\!p ) \left(\ft{2}{5-p}\right)^{p+2}
\bar{r}^{(3-p)^2 \over 2} \,ds\wedge dx_1 \wedge \cdots \wedge dx_{p+1} 
\quad+\, {\cal O}(L^{-1})
\end{eqnarray}
where $x_i=\rho^{(1)}\Omega^{(p)}_i$.
Thus the combination $\Omega$ for $p=0,2,4$ becomes
\begin{equation}
 \Omega= \ft{7-p}{8} \, \bar{r}^{3-p\over 4} L^{-1}\Gamma_{(p)} \quad +\,{\cal O}(L^{-2})
\end{equation}
where 
\begin{equation}
 \Gamma_{(p=0)}=\Gamma_{11} \Gamma^{\hat{s}\hat{x}_1},\quad
 \Gamma_{(p=2)}=-\Gamma^{\hat{s}\hat{x}_1 \hat{x}_2 \hat{x}_3},\quad
 \Gamma_{(p=4)}=-\Gamma_{11} \Gamma^{\hat{s}\hat{x}_1 \cdots \hat{x}_5}.
\end{equation}
Note that the local Lorentz indices here are those for the
Cartesian coordinates $\{ \hat{x}_i \}$ and
$\{ \hat{y}_l \}$, which are related to 
the angular coordinates $(\hat{\rho},\hat{\alpha}_a)$ and
$(\hat{\theta},\hat{\chi}_m)$ 
by
%$(\hat{\rho},\hat{\alpha}_a) \to \{ \hat{x}_i \}$ and 
%$(\hat{\theta},\hat{\chi}_m) \to \{ \hat{y}_l \}$ following 
%the corresponding space-time coordinate transformation 
$x_{i}=\rho^{(1)} \Omega_{i}^{(p)}$ and
$y_{l}=\theta^{(1)} \Omega_{l}^{(6-p)}$.
To obtain the expression for $p=4$, we have used the relation 
\begin{equation}
\frac{1}{4!} \Gamma^{\rho\sigma\lambda\tau}\, 
F_{\rho\sigma\lambda\tau}
= 
\frac{1}{6!}
\Gamma_{11} \Gamma^{\sigma_1\cdots \sigma_{6}}\, 
F_{\sigma_1\cdots \sigma_6}
\end{equation}
for ${}^{\ast}F_{(p+2)}=F_{(8-p)}$.

Finally, by redefining $\Theta$ as 
$$
 \Theta  \to 
\left(
\ft{2}{5-p} L \, \bar{r}^{3-p\over 4}
 \right)^{-{1\over 2}}  \Theta,
$$
the fermionic action at leading order becomes
\begin{eqnarray}
 S^{(2)}_{f,A}&=& {1 \over 2\pi} \int d\tau
\int^{2\pi\tilde{\alpha}}_{0} \! d\sigma \,
\Theta^{T}\Gamma_{\hat{s}}
\Bigg[ \sqrt{2} \Gamma_{+} \partial_{\tau}\Theta -i {7-p \over 2(5-p)}
\Gamma_{+} \Gamma_{(p)} \Gamma_{+} \Theta
\nonumber\\
&& \hspace{6cm} -i \sqrt{2}\,
%\left( {\cosh\tau \over \ell} \right)^{-(3-p)/(5-p)}
\bar{r}(\tau)^{-{3-p \over 2}}
\Gamma_{11} \Gamma_{+}
\partial_{\sigma} \Theta\Bigg]
\label{eq:SfA2}
\end{eqnarray}
where $\Gamma_{\pm}=(\Gamma_{\hat{\psi}}\pm \Gamma_{\hat{s}})/\sqrt{2}$.

We take the following representation of Gamma matrices: 
\begin{eqnarray}
&&\Gamma_{\hat{s}} = \pmatrix{0 & 1\cr -1& 0},\quad 
\Gamma_{\hat{\psi}} = \pmatrix{0 & 1\cr 1& 0},\quad 
\Gamma_{\hat{x}_i} = \pmatrix{\gamma_i & 0\cr 0& -\gamma_i} ,
\nonumber\\
&&
\Gamma_{\hat{y}_l} = \pmatrix{\gamma_{p+1+l} & 0\cr 0& -\gamma_{p+1+l}} ,\quad 
\Gamma_{11}=\pmatrix{\gamma_9&0\cr 0& -\gamma_9}\,
%&&
%\Gamma_+= \pmatrix{0& \sqrt2\cr 0& 0}\,,\;
%\Gamma_-= \pmatrix{0& 0\cr \sqrt2 & 0}\;
\label{eq:repGamma}
\end{eqnarray}
with $\gamma_9=\prod_{i=1}^8 \gamma_i$, 
and decompose $\Theta$ as
$$
\Theta={1\over \sqrt{2}}\pmatrix{\hat\theta \cr \theta}\, .
$$
Then the action becomes
\begin{equation}
S^{(2)}_{f,A} = - {1\over 2\pi}\int d\tau 
\int^{2\pi\tilde{\alpha}}_{0}
\! d\sigma \,
\theta^{T} 
\Big[\partial_{\tau} \theta
- i m_{f(p)} \gamma_{(p)}\theta
-i 
%\left( {\cosh\tau \over \ell} \right)^{-(3-p)/(5-p)}
\bar{r}(\tau)^{-{3-p \over 2}}
\gamma_{9}\partial_{\sigma}\theta\Big]
\label{eq:SfA2h}
\end{equation}
where 
%$\bar{r}(\tau)$ is given by (\ref{eq:defbarr}),
\begin{equation}
m_{f(p)}={7-p\over 2(5-p)}  
\end{equation}
and 
\begin{equation}
 \gamma_{(p=0)}= \gamma_{9}\gamma_{1},\quad
 \gamma_{(p=2)}= \gamma_{123},\quad
 \gamma_{(p=4)}= -\gamma_{9}\gamma_{12345}.
\end{equation}
As for the bosonic part, the lowest mode  
has constant mass in the present gauge.
% (\ref{eq:hgauge}). 

Note that only half of the components $\theta$ of the fermions 
appear in the action,
although we have not fixed the $\kappa$-symmetry. This is
related to the fact that the $\kappa$-symmetry at the leading 
order transforms only $\hat{\theta}$ which does not appear
in the action.
The action $S_{b}^{(2)}+S_{f}^{(2)}$ is not
world-sheet supersymmetric for $p=0,2,4$,
as discussed in Appendix~C.
In fact, masses of the bosons ($m_{x(p)}$, $m_{y(p)}$)
and the fermions ($m_{f(p)}$) do not match.

%%%
\subsection{type IIB}
%%%
For type IIB, the action quadratic in fermions 
after $\tau \to -i \tau$ is given by
\begin{equation}
S_{f,B} 
= {i\over 2\pi} \int\! d\tau d\sigma\, 
(\Theta^I)^{T}\Gamma_0
(\sqrt{h}h^{\alpha\beta}\delta^{IJ} +i \epsilon^{\alpha\beta} s_2^{IJ} )
\partial_{\alpha}X^{\mu} \Gamma_\mu  
{\cal D}_{\beta}^{JK} \Theta^K 
\end{equation}
where $\Theta^{I}$ ($I,J,K=1,2$) are two Majorana-Weyl spinors.
Derivative ${\cal D}_{\beta}$ is defined by 
\begin{equation}
{\cal D}^{IJ}_{\beta}\Theta^{J}=\nabla_{\beta}\Theta^{I}
+\partial_{\beta}X^{\mu}\Omega^{IJ} \Gamma_{\mu}\Theta^{J}
\end{equation}
where 
\begin{equation}
\Omega^{IJ} = - \ft18 e^\phi\, \Big(
%s^{IJ}_0\, \Gamma^\sigma\, \partial_\sigma\chi 
- \ft16 s^{IJ}_1\, \Gamma^{\mu_1\mu_2\mu_3}\, 
F_{\mu_1\mu_2\mu_3}\,
+ \ft1{240} 
s^{IJ}_0\, \Gamma^{\mu_1\cdots \mu_5}\, F_{\mu_1\cdots 
\mu_5}\Big) 
\end{equation}
for the background with RR three-form and five-form field strengths 
$F_{\mu_1\mu_2\mu_3}$ and $F_{\mu_1\cdots \mu_5}$. 
Matrices $s_{k}^{IJ}$ ($k=0,1,2$) are given by Pauli matrices as
$s_{0}=-i\sigma_{2}$, $s_{1}=\sigma_{1}$ and $s_2=\sigma_3$.

The action at the leading order is obtained following the same
procedure as for the type IIA case.
Substituting the background 
%(\ref{eq:metricDpdW}), 
(\ref{eq:vielbein}),
(\ref{eq:RRdW}) into the action, and expanding it 
around the geodesic (\ref{eq:trajG}), we obtain
\begin{eqnarray}
 S^{(2)}_{f,B}&=& {1 \over 2\pi} \int d\tau
\int^{2\pi\tilde{\alpha}}_{0} \! d\sigma \,
(\Theta^I)^{T} \Gamma_{\hat{s}}
\Bigg[ \sqrt{2} \Gamma_{+} \partial_{\tau}\Theta^I -i {7-p \over 2(5-p)}
\Gamma_{+} \Gamma_{(p)}^{IJ} \Gamma_{+} \Theta^J
\nonumber\\
&& \hspace{6cm} +i \sqrt{2}\,
%\left( {\cosh\tau \over \ell} \right)^{-(3-p)/(5-p)}
\bar{r}(\tau)^{-{3-p \over 2}}
s_2^{IJ} \Gamma_{+}
\partial_{\sigma} \Theta^J  \Bigg]
\label{eq:SfB2}
\end{eqnarray}
where
\begin{equation}
\Gamma_{(p=1)}^{IJ} = 
- s_1^{IJ} \Gamma^{\hat{s}\hat{x}_1\hat{x}_2 },\quad
 \Gamma_{(p=3)}^{IJ}
= s_{0}^{IJ} \Gamma^{\hat{s}\hat{x}_1 \hat{x}_2 \hat{x}_3 \hat{x}_4}.
\end{equation}
By choosing the representation of Gamma matrices (\ref{eq:repGamma}), 
the action becomes
\begin{equation}
S^{(2)}_{f,B} =  -{1\over 2\pi}\int d\tau 
\int^{2\pi\tilde{\alpha}}_{0}
\! d\sigma \,
(\theta^I)^{T} 
\Big[\partial_{\tau} \theta^I
- i m_{f(p)} \gamma_{(p)}^{IJ}\theta^J
+i 
%\left( {\cosh\tau \over \ell} \right)^{-(3-p)/(5-p)}
\bar{r}(\tau)^{-{3-p \over 2}}
s_2^{IJ}  \partial_{\sigma}\theta^J \Big]
\label{eq:SfB2h}
\end{equation}
with
\begin{equation}
 \gamma_{(p=1)}= s_1 \gamma_{12},\quad
 \gamma_{(p=3)}= - s_0 \gamma_{1234}.
\end{equation}
As in the case of type IIA, only $\theta$ appears in the action.
For $p=3$, bosonic and
fermionic masses of the lowest modes coincide 
($ m_{x(p=3)} = m_{y(p=3)} = m_{f(p=3)}=1 $ ),
because of the presence of the world-sheet supersymmetry.

\section{Amplitudes and two-point functions for the fermionic sector}

Let us study amplitudes for the fermionic sector of the
superstrings, and obtain two-point functions of 
fermionic operators of the gauge theories.
We first calculate the amplitude for the supergravity mode in 
section 4.1, 
and following the procedure similar to the bosonic case,
we interpret them as two-point functions.
In section 4.2, we examine consistency of this result with 
the one previously obtained in \cite{SeYo, Se} through 
the supergravity analysis. We discuss quantization of the
string higher modes in section 4.3.

\subsection{S-matrix for the supergravity modes}
The action of the fermionic fluctuations for type IIA (\ref{eq:SfA2h}) 
and type IIB (\ref{eq:SfB2h}) can be written in a unified manner as
\begin{eqnarray}
S^{(2)}_{f} &=&  -{1\over 2\pi}\int d\tau 
\int^{2\pi \tilde{\alpha} }_{0}
\! d\sigma \, \Big[
\theta_+^{T} \partial_{\tau} \theta_+ 
+ \theta_-^{T} \partial_{\tau} \theta_-
\nonumber\\
&& \hspace*{2cm} - 2 i m_{f(p)} \theta_+^{T} \gamma_{(p)} \theta_-
-i \bar{r}(\tau)^{-{3-p \over 2}}
( \theta_+^T \partial_{\sigma} \theta_+ -  \theta_-^T \partial_{\sigma} \theta_- )
\big]
\end{eqnarray}
where $\tilde{\alpha}={5-p\over 2}{J\over L^2} $ and
\begin{eqnarray}
&& \theta=\theta_+ + \theta_- ,\quad \gamma_9 \theta_{\pm} = \pm \theta_{\pm} 
 \hspace*{2.45cm}  \mbox{for IIA},
\\
&& (\theta^1, \theta^2) = (\theta_-, \theta_+) \quad \mbox{or}\quad 
 s_2 \theta_{\pm} = \mp \theta_{\pm} 
 \qquad  \mbox{for IIB}.
\end{eqnarray}
By expanding $\theta_{\pm}$ into Fourier modes
\begin{equation}
\theta_{\pm}= {1\over \sqrt{\tilde{\alpha}}}\,
\sum_{n=-\infty}^{\infty}
\theta_{\pm,n}(\tau) e^{i {n\over \tilde{\alpha}} \sigma}
,
\end{equation}
the action is rewritten as
\begin{eqnarray}
S^{(2)}_{f} &=&  -\sum_{n=-\infty}^{\infty} \int d\tau\, 
\Bigg[
\theta_{+,-n}^{T} \partial_{\tau} \theta_{+,n} 
+ \theta_{-,-n}^{T} \partial_{\tau} \theta_{-,n}
- 2 i m_{f(p)} \theta_{+,-n}^{T} \gamma_{(p)} \theta_{-,n}
\nonumber\\
&& \hspace*{2cm} 
+ \,{n\over \tilde{\alpha}} \, \bar{r}(\tau)^{-{3-p \over 2}}
( \theta_{+,-n}^T \theta_{+,n} -  \theta_{-,-n}^T  \theta_{-,n} )
\bigg].
\end{eqnarray}
Equations of motion are given by
\begin{equation}
\partial_{\tau} \theta_{\pm,n} 
\pm {n\over \tilde{\alpha}} \, \bar{r}(\tau)^{-{3-p \over 2}} \theta_{\pm,n} 
-  i m_{f(p)} \gamma_{(p)} \theta_{\mp,n}=0.
\end{equation}

Canonical quantization is performed in the standard way.
Momentum for $\theta_{\pm,n}$ is 
\begin{equation}
 P_{\pm,n} \equiv 
i {\partial\over \partial \dot{\theta}_{\pm,n}} {\cal L}
= i\theta_{\pm,-n}\,,
\end{equation}
and we have the corresponding second-class constraints
\begin{equation}
 P_{\pm,n} -i \theta_{\pm,-n}\approx 0 .
\end{equation}
Dirac brackets for $\theta_{\pm,n}$ are calculated as 
\begin{equation}
\{\theta_{s,n}, \theta_{s',n'} \}_{\rm DB} = -{i\over 2} \delta_{s,s'} \delta_{n,-n'}
\end{equation}
where $s, s' =\pm$. 
Then the canonical anti-commutation relations are
\begin{equation}
\{\theta_{s,n}, \theta_{s',n'} \} = {1\over 2} \delta_{s,s'} \delta_{n,-n'} .
\label{eq:anticomm}
\end{equation}
Hamiltonian of the system is
\begin{equation}
H =   \sum_{n=-\infty}^{\infty} \bigg[
- 2 i m_{f(p)} \theta_{+,-n}^{T} \gamma_{(p)} \theta_{-,n}
+ {n\over \tilde{\alpha}} \, \bar{r}(\tau)^{-{3-p \over 2}}
( \theta_{+,-n}^T \theta_{+,n} -  \theta_{-,-n}^T  \theta_{-,n} )
\bigg].
\end{equation}
Since $\theta$ is real, $\theta_{\pm,n}$ must satisfy 
$\theta_{\pm,n}(\tau)^\dagger = \theta_{\pm, -n}(-\tau)$.
Note that the complex conjugation is accompanied by the
reflection of $\tau$ in the Euclidean formulation~\cite{AsSeYo}.
%Note that $\theta_{\pm,n}$ satisfy the reflection condition 
%$\theta_{\pm,n}(\tau)^\dagger = \theta_{\pm, n}(-\tau)$
%instead of the reality condition for the Minkowski case.

Let us concentrate on the supergravity modes $n=0$. 
The equations of motion for $\theta_{\pm,0}$ are 
\begin{equation}
\partial_{\tau} \theta_{\pm,0} 
-  i m_{f(p)} \gamma_{(p)} \theta_{\mp,0}=0.
\end{equation}
This leads to the mode expansion:
\begin{eqnarray}
\theta_{-,0}^\alpha &=&
f_0^{(+)} d_0^\alpha + f_0^{(-)} d_0^{\alpha \dagger},
\\
\theta_{+,0}^{\alpha} &=&
-i f_0^{(+)} \gamma_{(p)}^{\alpha\beta}d_0^\beta ,
+ i f_0^{(-)} \gamma_{(p)}^{\alpha\beta}d_0^{\beta \dagger}
\label{eq:sugramodeexp}
\end{eqnarray}
with 
\begin{equation}
f_0^{(+)} ={1\over 2} e^{-m_{f(p)} \tau} ,\quad
f_0^{(-)} ={1\over 2} e^{m_{f(p)} \tau}
\end{equation}
and 
\begin{equation}
\{d_0^{\alpha} , d_0^{\beta \dagger} \} = \delta_{\alpha,\beta}.
\end{equation}
Here, we have written explicitly the indices
$\alpha,\beta=1,\ldots,8$ for
the SO(8) spinors
in the subspace of negative $\gamma_{9}$ (positive $s_{2}$)
eigenvalue for type IIA (type IIB).
Note that (\ref{eq:sugramodeexp}) is consistent with
the anti-commutation relation (\ref{eq:anticomm})
and the reality condition.
%Note that we have chosen the normalization of 
%$f^{(\pm)}_0$ so that they are consistent 
%with the anti-commutation relation (\ref{eq:anticomm}). 

Hamiltonian for the $n=0$ sector is
\begin{eqnarray}
H_0  &=&   - 2 i m_{f(p)} \theta_{+,0}^{T} \gamma_{(p)} \theta_{-,0}
\nonumber\\
&=& m_{f(p)} \left(d_0^{\alpha \dagger} d_0^{\alpha} 
- {1\over 2}\delta_{\alpha,\alpha}\right).
\end{eqnarray}
Thus the S-matrix $S_{f}(T)$ from $\tau=-T$ to $\tau=T$ is 
\begin{eqnarray}
S_{f}(T) &=& \exp \left( -\int^T_{-T} d\tau H \right) 
\nonumber\\
&=& \exp \left[-  m_{f(p)} 
\big(2 d_0^{\alpha \dagger} d_0^{\alpha} 
- \delta_{\alpha,\alpha}\big)
T \right]. 
\end{eqnarray}
As for the bosonic excitations, this is rewritten in the
$T\to \infty$ limit as 
\begin{equation}
  S_{f}(T) \sim \left( |t_f-t_i|\Lambda
 \right)^{-{7-p \over{2(5-p)}} 
(2 d_0^{\alpha \dagger} d_0^{\alpha} 
- 8)
}.  
\label{eq:fermiamp}
\end{equation}
%for $T\to \infty $.
Thus, when $N_{f}$ fermionic oscillators of the supergravity modes
are excited, the amplitude
including the classical contribution becomes
\begin{equation}
 S(T)\sim \left( |t_f-t_i|\Lambda
 \right)^{-{4\over 5-p}(J+{1\over 2}N_{f})-{7-p \over(5-p)}N_{f}-2c_{0}}. 
\label{eq:fermismat}
\end{equation}
Here, we have the term $N_{f}/2$ in the power, 
due to the spin angular momentum of fermionic oscillators,
%\footnote{Under the rotation 
%(boost in the Euclidean formulation)
%$\psi\to\psi+\Delta$, wave function $e^{-J\psi}$
%of the ground state with angular momentum $J$, 
%transforms into $e^{-J\Delta}e^{-J\psi}$. This rotation, 
%which is assumed to be in the $s$-$\psi$ plane, 
%acts on spinors as  $\Theta\to
%\exp ({\Delta\over 2} \Gamma_{\hat{s}\hat{\psi}})\Theta$.
%Physical fermions satisfy 
%$\Gamma_{\hat{s}\hat{\psi}}\Theta=-\Theta$ as
%we have seen in section 3. A spinor creation operator
%gains a factor 
%$\exp ({\Delta\over 2} \Gamma_{\hat{s}\hat{\psi}})=
%\exp(-{\Delta\over 2})$ under the rotation, so 
%the angular momentum increases by $1/2$
%as we excite one fermionic oscillator.}, 
as we have discussed in section 2.2.

\subsection{Consistency with the supergravity analysis}

First, we calculate the zero-point energy $c_{0}$ from the supergravity
modes, by summing up the contributions from $8=(p+1)+(7-p)$ bosonic  
and 8 fermionic oscillators%
\footnote{We leave the evaluation of zero-point energies of the
string higher modes $(n\ne 0)$ for future work.}:
\begin{equation}
 c_{0}={1\over 2}\left((p+1)\cdot 1 +(7-p)\cdot {2\over 5-p}
-8\cdot {7-p\over 2(5-p)}\right)=-{(3-p)^{2}\over 2(5-p)}.
\end{equation} 
This does not vanish except for $p=3$ where the theory
is world-sheet supersymmetric.

Taking  this $c_{0}$ into account, 
two-point function for the ground state operator
${\cal O}\equiv{\rm Tr} Z^{J}$ discussed in section 2.4
becomes
\begin{equation}
 \langle \bar{{\cal O}}(t_f){\cal O}(t_i)\rangle
\sim (|t_f-t_i|\Lambda)^{-{4\over 5-p}J+{(3-p)^{2}\over 5-p}}.
\label{eq:ostring}
\end{equation}
Let us consider the case when one fermionic supergravity
mode is excited. As a natural extension of the BMN's operator correspondence,
we assume that the operator corresponding to this supergravity
state is ${\cal O}_{\chi}\equiv{\rm Tr}(\chi Z^{J})$, where 
$\chi$ is half of the 16 spinors
of the $(p+1)$-dimensional SYM. From (\ref{eq:fermismat}),
the two-point function of ${\cal O}_{\chi}$ is
\begin{equation}
 \langle \bar{{\cal O}}_{\chi}(t_f){\cal O}_{\chi}(t_i)\rangle
\sim (|t_{f}-t_{i}|\Lambda)^{-{4\over 5-p}(J+{1\over 2})-{7-p\over 5-p}
+{(3-p)^{2}\over 5-p}}. 
\label{eq:ochistring}
\end{equation}

These are consistent with the results of~\cite{SeYo, Se}
where the correlator for $p=0$ was obtained by the 
supergravity analysis following the
GKP/W prescription. We shall briefly summarize the
relevant results of~\cite{SeYo, Se}. Apart from constant
factors, the two-point functions of bosonic operators~\cite{SeYo} 
are of the form
 \begin{equation}
  \langle \bar{{\cal O}}_{b}(t_f){\cal O}_{b}(t_i)\rangle
\sim |t_{f}-t_{i}|^{-{4\over 5}J-{14\over 5}n-1},
\label{eq:obsugra}
 \end{equation}
and those for fermionic operators~\cite{Se} are
\begin{equation}
  \langle \bar{{\cal O}}_{f}(t_f){\cal O}_{f}(t_i)\rangle
\sim |t_{f}-t_{i}|^{-{4\over 5}J-{14\over 5}n-{7\over 5}}.
\label{eq:ofsugra}
\end{equation}
where ${\cal O}_{b}$ are conserved currents (energy-momentum tensor)
and ${\cal O}_{f}$ are the supercurrents of Matrix theory~\cite{TaRa}.
Here, $n$ is specified by 11-dimensional 
tensor structure of the operators when 
interpreted as DLCQ M-theory;
$n=1-n_{+}+n_{-}+{1\over 2}s$ where $n_{\pm}$ are the 
number of the $\pm$ indices, and we have  
$s=+1$ ($s=-1$) for the so-called dynamical supercurrents $q^{M}$,
(kinematical supercurrents $\tilde{q}^{M}$). 

The ground state operator ${\cal O}$ has $n=-1$ because it
belongs to
\begin{equation}
 T^{++}_{J}\sim {\rm STr}(X_{I_1}\cdots X_{I_J}+\cdots)
\end{equation}
where $I_{1},I_{2},\cdots=1,\ldots 9$.
We see that the supergravity result (\ref{eq:obsugra})
with $n=-1$ indeed agrees with the result from string theory
(\ref{eq:ostring}). Note that the $J$ independent part was 
correctly reproduced by the contribution of 
the zero-point energy $c_{0}$.
The fermionic operator ${\cal O}_{\chi}$ belongs to
\begin{equation}
\tilde{q}^{+}_{J}\sim  {\rm STr}(\chi
X_{i_1}\cdots X_{i_J}+\cdots), 
\end{equation}
and has $n=-1/2$.
The supergravity result (\ref{eq:ofsugra}) 
also agrees with the string result (\ref{eq:ochistring}).

For generic supergravity states constructed by acting the
bosonic creation operators on the above states (the ground state
and the lowest fermionic states), the string and the
supergravity analysis give the same result. Indeed, 
as mentioned in section 2, excitations of bosonic oscillators
with mass $m_{x(p=0)}=1$ correspond to insertions of
$D_{0}$ into the trace, and those with $m_{y(p=0)}=2/5$ 
correspond to the $X_{l}$ insertions. The $D_{0}$ insertion
changes the index structure of an operator by
$n\to n+1$ and $J\to J-1$~\cite{SeYo, Se, AsSeYo},
which leads to the multiplication 
of the supergravity result (\ref{eq:obsugra}), (\ref{eq:ofsugra}) 
by a factor $|t_{f}-t_{i}|^{-2}$. This is what we obtain
by exciting an  oscillator with $m_{x(p=0)}$ in the string picture.
Similarly, the $X_{l}$ insertion amounts to $J\to J+1$, and
we have $|t_{f}-t_{i}|^{-4/5}$ factor from the supergravity analysis,
in accord with the string result.

We also find that our result is consistent with available
partial results of the supergravity analysis 
for other D$p$-branes%
\footnote{We thank T.~Gherghetta for drawing our attention
to this issue.}~\cite{AnHaLuPi, GhOz}. By considering
a particular mode of the graviton on the near-horizon D$p$-brane
background, it has been found that the two-point function
of energy-momentum tensor of the $(p+1)$-dimensional SYM
is of the form~\cite{GhOz}
\begin{equation}
 \langle T(t_{f})T(t_{i})\rangle
\sim {1\over |t_{f}-t_{i}|^{2\alpha+p+1}},\qquad \alpha={7-p\over 5-p}.
\label{eq:ttsugra}
\end{equation}
In our approach%
\footnote{Our method is valid for large $J$, but here 
we shall apply the result to the case of small $J$.}, 
energy-momentum tensor of the form 
${\rm Tr}(D_{i}ZD_{j}Z)$ corresponds to the state constructed
by applying two creation operators with mass $m_{x(p)}=1$
on the ground state with $J=2$. We can easily see that
the two-point function (\ref{eq:twopointb}) equals (\ref{eq:ttsugra}).
For the other components of the energy-momentum tensor,
${\rm Tr}(D_{i}ZD_{j}\phi_{l})$, ${\rm Tr}(D_{i}\phi_{l_1} D_{j}\phi_{l_2})$,
we have the same answer.
%
%
%Thus, the two-point function is given by substituting 
%$J=2$, $N_{0}^{x}=2$, $N_{0}^{y}=0$, $c_{0}=-(3-p)^{2}/(5-p)$
%into (\ref{eq:twopointb}); this  equals (\ref{eq:ttsugra}).

Finally, note that for $p=3$,
the two-point function of ${\cal O}_{\chi}$ from
(\ref{eq:ochistring}) is
$|t_{f}-t_{i}|^{-2\Delta}$ with $\Delta=J+3/2$,
as it should. (${\cal O}_{f}$ is BPS and has no anomalous dimension.)

\subsection{Quantization of the string higher modes}
%%% sect 4.3. %%%
% \subsection{Effect of string higher modes}

We return to the problem of quantization of $\theta_{\pm,n}(\tau)$ 
including string higher modes $n\ne 0$.
Equations of motion for $\theta_{\pm,n}$ are rewritten as
\begin{eqnarray}
&& \partial_{\tau} \tilde{\theta}_{+,n} 
+ {n\over \tilde{\alpha}} \, \bar{r}(\tau)^{-{3-p \over 2}} \tilde{\theta}_{+,n} 
-  m_{f(p)} \theta_{-,n}=0,
\nonumber\\
&& \partial_{\tau} \theta_{-,n} 
- {n\over \tilde{\alpha}} \, \bar{r}(\tau)^{-{3-p \over 2}} \theta_{-,n} 
-  m_{f(p)}  \tilde{\theta}_{+,n}=0
\label{eq:EOMnt}
\end{eqnarray}
where $ \tilde{\theta}_{+,n}=  i \gamma_{(p)} \theta_{+,n}$.
(Note that $\gamma_{(p)}^T=\gamma_{(p)}^{-1}=-\gamma_{(p)}$.)
General solution for the pair
$(\theta_{-}^n, \tilde{\theta}_{+}^n)$ is given as a linear combination of 
two independent solutions
$(\phi_n^{(1)}, \psi_n^{(1)})$ and $(\phi_n^{(2)}, \psi_n^{(2)})$. 
For an arbitrary set of solutions, 
we have a relation 
\begin{equation}
{d \over d\tau} \left(
\phi_n^{(1)} \psi_n^{(2)} - \phi_n^{(2)} \psi_n^{(1)} 
\right) =0 .
\end{equation}

We express the general solution by spinor operators $d_n$ and $d_n^\dagger$ as
\begin{equation}
\left(\theta_{-}^n (\tau) , \tilde{\theta}_{+}^n(\tau) \right)
= \left( \phi_n^{(1)}(\tau) \,d_n+  \phi_n^{(2)}(\tau)  \,d_n^\dagger \;,\;  
\psi_n^{(1)}(\tau) \, d_n+  \psi_n^{(2)}(\tau)\,  d_n^\dagger  \right).
\end{equation}
We can set $\phi_{-n}^{(1)} (-\tau) = \phi_n^{(2)}(\tau)$ from the time reflection 
symmetry $\theta_{\pm,n}(\tau)^\dagger = \theta_{\pm, -n}(-\tau)$. 
We choose 
\begin{equation}
\phi_n^{(1)} \psi_n^{(2)} - \phi_n^{(2)} \psi_n^{(1)} = {1\over 2}
\label{eq:solcond1}
\end{equation}
and 
\begin{equation}
\phi_n^{(1)} =- \psi_{-n}^{(1)},\quad  \phi_n^{(2)} = \psi_{-n}^{(2)} .
\label{eq:solcond2}
\end{equation}
This is possible since 
$(\psi_{-n}, \phi_{-n})$ is a solution for $(\theta_{-}^n, \tilde{\theta}_{+}^n)$ 
if $(\phi_{-n}, \psi_{-n})$ is a solution for 
$(\theta_{-}^{-n}, \tilde{\theta}_{+}^{-n})$.
Then the quantization condition (\ref{eq:anticomm})
is consistently satisfied with 
\begin{equation}
\{d_n, d_{n'}^{\dagger} \} = \delta_{n,-n'}.
\end{equation}
The Hamiltonian is written as 
\begin{eqnarray}
H &=& \sum_{n=-\infty}^{\infty} \bigg[
 2  m_{f(p)} \tilde{\theta}_{+,-n}^{T} \theta_{-,n}
- {n\over \tilde{\alpha}} \, \bar{r}^{{p-3 \over 2}}
( \tilde{\theta}_{+,-n}^T \tilde{\theta}_{+,n} +  \theta_{-,-n}^T  \theta_{-,n} )
\bigg]
\nonumber\\
&=& \sum_{n=-\infty}^{\infty} \sum_{\alpha=1}^8 
 \Bigg\{
- 2 \bigg[ m_{f(p)} \Big( \phi_{n}^{(1)} \Big)^2
+{n\over \tilde{\alpha}} \bar{r}^{{p-3 \over 2}} 
\phi_{n}^{(1)} \phi_{-n}^{(1)} 
 \bigg] d_{-n}^{\alpha} d_n^\alpha  
\nonumber\\
& &  + 2 \bigg[ m_{f(p)} \Big( \phi_{n}^{(1)} \phi_{n}^{(2)} + \phi_{-n}^{(1)} \phi_{-n}^{(2)} \Big)
+{n\over \tilde{\alpha}} \bar{r}^{{p-3\over 2}} 
\Big(\phi_{-n}^{(1)} \phi_{n}^{(2)} - \phi_{n}^{(1)} \phi_{-n}^{(2)} \Big)
 \bigg] \bigg( d_{-n}^{\alpha\dagger} d_n^\alpha-{1\over 2}  \bigg) 
\nonumber\\
&& + 2 \bigg[ m_{f(p)} \Big( \phi_{n}^{(2)} \Big)^2
-{n\over \tilde{\alpha}} \bar{r}^{{p-3 \over 2}} 
\phi_{n}^{(2)} \phi_{-n}^{(2)} 
 \bigg] d_{-n}^{\alpha\dagger} d_n^{\alpha\dagger}  
\Bigg\}.
\end{eqnarray}

In general, this Hamiltonian is time-dependent $H=H(\tau)$.
S-matrix is defined by the integration of the anti-time 
ordered product
\begin{equation}
 S_{f}(T)={\cal T}_{-} \exp\left[ -\int^{T}_{-T} d\tau H(\tau)\right].  
\end{equation}
A method proposed in~\cite{AsSeYo} for dealing with such
time-dependent oscillators  
is to first obtain the S-matrix using the above $d_{n}^{\alpha}$,
$d_{-n}^{\alpha}$ basis
and then perform Bogoliubov transformation to define
oscillators which diagonalize the S-matrix. 
We defer further discussion on the higher modes 
to a separate paper~\cite{AsSe}, in which the behavior
of correlators corresponding to the bosonic and fermionic
string higher excitations will be studied
in detail.

We finish this section by presenting the S-matrix for the simple
case of $p=3$ where mass is constant $m_{(p=3)}=1$. 
The equations of motion (\ref{eq:EOMnt}) are exactly solved as 
\begin{equation}
\theta_{-}^n (\tau) = 
c_{\pm}e^{\mp \sqrt{1+{n^2\over  \tilde{\alpha}^2} }\, \tau},
\quad
\tilde{\theta}_{+}^n(\tau) 
=
c_{\pm}
\left(- {n\over  \tilde{\alpha}}  \mp \sqrt{1+{n^2\over  \tilde{\alpha}^2}}
\right)
e^{\mp \sqrt{1+{n^2\over  \tilde{\alpha}^2} }\, \tau} .
\end{equation}
To satisfy (\ref{eq:solcond1}), we set 
\begin{equation}
\phi_{n\ge 0}^{(1)} =  
{1\over 2}\left(1+{n^2\over  \tilde{\alpha}^2} \right)^{-1/4}
e^{- \sqrt{ 1+{n^2\over  \tilde{\alpha}^2} } \, \tau} ,
\quad 
\phi_{n\ge 0}^{(2)} =  
{1\over 2}\left(1+{n^2\over  \tilde{\alpha}^2} \right)^{-1/4}
e^{+ \sqrt{1+{n^2\over  \tilde{\alpha}^2} } \, \tau} .
\end{equation}
The remaining $\phi_{n < 0}^{(1)}$ and $\phi_{n < 0}^{(2)}$ 
are determined by (\ref{eq:solcond2}).
Then the Hamiltonian becomes 
\begin{equation}
 H_f^{p=3} = \sum_{n=-\infty}^{\infty} \sum_{\alpha=1}^8
\sqrt{ 1+{n^2\over  \tilde{\alpha}^2} } \,
\bigg( d_{n}^{\alpha\dagger} d_{-n}^\alpha-{1\over 2}  \bigg),
\end{equation}
and the S-matrix is readily obtained as
\begin{equation}
  S_{f}(T)^{p=3} = 
\prod_{n=-\infty}^{\infty}
\exp \bigg[ -2 \sqrt{ 1+{n^2\over  \tilde{\alpha}^2} } 
\bigg( d_{n}^{\alpha\dagger} d_{-n}^\alpha - 4  \bigg) 
T  
\bigg] .
\end{equation}

\section{Conclusions}
We have investigated the holographic correspondence
for D$p$-branes with $0\le p \le 4$, following an 
approach proposed in \cite{DoShYo, AsSeYo} which is based
on  Euclidean strings on the double Wick rotated 
D$p$-brane backgrounds. In the first part of the paper, we have
reviewed our basic strategy and the results of~\cite{AsSeYo}: 
we saw that classical
trajectory of point-particle (tunneling null geodesic) connects
two points on the boundary of the $(p+2)$-dimensional AdS-like
space; we obtained the amplitudes for the bosonic fluctuations 
in the semi-classical limit around tunneling null geodesic; 
we interpreted the amplitudes
as two-point functions of gauge theory, with an identification
of the gauge theory operators which is a natural extension of 
the BMN proposal~\cite{BeMaNa}. 

Then, we  computed the amplitudes for the fermionic sector. 
We noted that the superstring action in 
the semi-classical limit does not have world-sheet supersymmetry
when $p\ne 3$; energy of bosonic and fermionic modes do not match.
We obtained the two-point functions of the operators 
corresponding to the supergravity modes which are power-law
behaved. The result is consistent with 
the one previously obtained in~\cite{SeYo, Se} by applying 
the prescription of Gubser, Klebanov and Polyakov~\cite{GuKlPo}
and Witten~\cite{Wi} to the D0-brane background.
This agreement with the supergravity analysis
is true not only at the leading order in $J$ but 
also to the subleading order, 
as we take into account the zero-point energies
from bosonic and fermionic fluctuations.
We have also seen that our result is consistent with
available partial results~\cite{AnHaLuPi, GhOz}
of supergravity analyses for other D$p$-branes.

We have discussed string higher modes only 
briefly in this paper. We have obtained the Hamiltonian
for the fermionic fluctuations which are time-dependent
for $p\ne 3$, but we postpone the analysis of the amplitude
to a separate paper~\cite{AsSe}. With the method developed
in~\cite{AsSeYo}, we can compute the amplitude and obtain
the two-point functions in principle. In particular, 
we can perform a perturbative expansion of these amplitudes
taking the solutions for the supergravity modes as basis functions, 
and see the deviation from the power-law correlators.
Analysis along this line will be reported in~\cite{AsSe}.

We should mention that we have no results at present
from the gauge theory computations for the correlators of
non-conformal SYM theories, which would test our prediction 
from holography. 
Low dimensional SYM have severe infra-red divergences, and the
correlators in the strong coupling region will not be
given by a simple perturbation theory.
We hope our analysis provides a hint for future studies
of those SYM theories. One of our predictions of this 
and the previous~\cite{AsSeYo} paper is that certain 
operators (Matrix theory currents) which are analogs 
of chiral primary operators of $D$=4, ${\cal N}=4$ SYM
have power-law correlators, while other operators 
(which correspond to string higher modes) do not. 
This seems to suggest that 
supersymmetric protection is working for the
former operators, more or less similarly to the case of
the chiral primary operators whose scaling dimensions
are protected by the superconformal algebra. 
Supersymmetric version
of the `generalized conformal symmetry'~\cite{JeYo} 
possibly plays a role.

It would be 
interesting to study the dynamics of the string further in 
the double Wick rotated framework.
There are a few directions for future studies
even for the $p=3$ case.
An immediate problem is how to understand the
works of Tseytlin and others~\cite{Ts} on the 
correspondence between macroscopic strings 
and non-BPS operators in our approach. 
Detailed study of the three and higher point functions, 
whose qualitative feature was discussed in~\cite{DoShYo},
is also important. This would teach us how to describe general
bulk physics in terms of the boundary theory.

Finally, we can apply our method to more general
examples of the holographic correspondence, 
such as the one 
between the AdS-Schwarzschild black hole and the finite
temperature $D$=4 SYM~\cite{Wi2}. 
In~\cite{FiHuKlSh}, the authors studied the analytic
structure of correlators computed by the geodesic 
approximation, and clarified the relation between
the correlators in Minkowski theory and those 
in Euclidean theory. 
They argued that information
behind the horizon is encoded in boundary gauge theory.
It would be interesting to study the analytic structure
of the correlators obtained in our approach, especially,
of those corresponding to string higher excitations.

\bigskip
%%%%%%%%%%%%%%%%%%%%%%%%%%%%%%%%%%%%
\noindent
Acknowledgments

We would like to thank T.~Yoneya for valuable discussions and comments.
Y.~S. is also grateful for A.~Jevicki for discussions.
The work of M.~A. is supported in part by the Grants-in-Aid for 
the 21st Century COE ``Center for Diversity and Universality in Physics'' 
from the Ministry of Education, Culture, Sports, 
Science and Technology (MEXT) of Japan.
The work of Y.~S. is supported in part by the Research Fellowships
of the Japan Society for the Promotion of Science for Young Scientists.

\bigskip

%%%%%%%%%%%%%%%%%%%%%%%%%%%%%%%%%%%%

\section*{Appendix}

\appendix 
%%%%%%
\section{Superstring action in the conformal gauge}
\setcounter{equation}{0}
\renewcommand{\theequation}{\Alph{section}.\arabic{equation}}
%%%

In this appendix, we give a brief discussion on the action for the 
fluctuation in the conformal gauge. We review the bosonic part of the
superstring action which was studied in~\cite{AsSeYo}, and also obtain the fermionic
part. We will see the equivalence of the superstring action
in the conformal gauge and the one in the gauge condition adopted in 
the text.

The bosonic part of the superstring action after double Wick rotation 
is given by 
\begin{equation}
S_{b}={1\over 4\pi}\int d\tau_c \int_{0}^{2\pi} d\sigma_c
\sqrt{h} h^{\alpha\beta} \partial_{\alpha}x^{\mu}
\partial_{\beta}x^{\nu}\tilde{g}_{\mu\nu}
\label{eq:actionconpre}
\end{equation}
with 
\begin{equation}
d\tilde{s}^2= L^{2}\left\{ r^{7-p\over 2} (dt^{2}+d x_{a}^{2})
+ r^{-{7-p \over 2}} 
(dr^{2}+r^{2}(d\theta^2- \cos^2\theta d\psi^2 + \sin^2\theta  
d\Omega_{6-p}^{2}))\right\}.
\end{equation}
We fix the world-sheet metric by conformal gauge 
$\sqrt{h} h^{\alpha\beta} =\delta^{\alpha\beta}$.

The tunneling null geodesic is given as a classical solution 
of this action satisfying $\theta=x_a=0$. 
We can identify two conserved charges for the solution
\begin{equation}
  E\equiv  L^{2} r^{7-p\over 2} \dot{t}\,, 
\quad J\equiv L^{2} r^{-{3-p \over 2}}\dot{\psi}
\end{equation}
where $E$ represents the energy and $J$ the angular momentum along $\psi$.
After rescaling $(E/L^2)\tau_c \to \tau_c $ and defining $\ell\equiv J/E$,
the geodesic is represented as
\begin{equation}
\dot{r}=
\left\{
\begin{array}{@{\,}ll}
\sqrt{ \ell^2 r^{5-p}-1} & (\tau_c \ge 0)
\\
-\sqrt{ \ell^2 r^{5-p}-1} & (\tau_c \le 0)
\end{array}
\right.\,,
\label{eq:defrtau}
\end{equation}
with   
$r(\tau_c=0)=0$, 
$\dot{t}= r^{-{7-p\over 2}} $ and $ \dot{\psi}= \ell r^{3-p \over 2}$.
The profile of this geodesic in the $(t,r)$ plane is 
\begin{equation}
t^2 +z^2= 
\tilde{\ell}^2\, , \quad
z\equiv \ft{2}{5-p} r^{-{5-p\over 2}}\, ,\quad
\tilde{\ell}\equiv {2\over 5-p}\ell.
\end{equation}
We can write $\psi(\tau_c)$ explicitly in terms of $r(\tau_c)$ as
\begin{equation}
\psi(\tau_c)= \ft{2}{5-p} \ln \left(
\ell r(\tau_c)^{5-p \over 2} 
\pm \sqrt{\ell^2 r(\tau_c)^{5-p} -1}
\right).
\end{equation}
We perform the semi-classical expansion of the 
action (\ref{eq:actionconpre}) 
around this geodesic as in section~\ref{sec:2.3},  
and obtain 
\begin{equation}
S_b^{(2,c)}={1\over 4 \pi}
\int d\tau_c \int_0^{2\pi E/L^2} \!\!\! d\sigma_c\,
( \dot{X_i}^2 +{X'_i}^2 + \dot{Y_l}^2 +{Y'_l}^2
+m_X^2 X_i^2 + m_Y^2 Y_l^2) 
\label{eq:actionbosconf}
\end{equation}
where we have time-dependent masses
\begin{eqnarray}
m_X^{2} &=& -
{(7-p) \over 16 \bar{r}^{2}}
[ (3-p) +(3p-13)\ell^{2} \bar{r}^{5-p}  ]
,
\label{eq:timedepMX}
\\
m_Y^{2} &=& - {(7-p) \over 16 \bar{r}^{2}}
[ (3-p) -(p+1)\ell^{2} \bar{r}^{5-p}  ].
\label{eq:timedepMY}
\end{eqnarray}
Here, $i=1,\cdots,p+1$, $l=1,\cdots,7-p$, and
$\bar{r}=r(\tau_c)$ is defined by (\ref{eq:defrtau}).

The action (\ref{eq:actionbosconf}) is transformed to the
one used in the text (\ref{eq:actionbos2}), by making the
field redefinitions
\begin{equation}
X_i = \ft{2}{5-p} \bar{r}^{-{3-p\over 4}} x_i,
\quad
Y_l = \bar{r}^{-{3-p\over 4}} y_l,
\end{equation}
and the reparametrization
$(\tau_c,\sigma_c)\rightarrow (\tau,\sigma)$ with
\begin{equation}
\sigma = \ft{5-p}{2}\ell \sigma_c \,, \quad 
{d\tau \over d\tau_c} = \ft{5-p}{2}\ell \bar{r}^{3-p\over 2}\,. 
\label{eq:wsredef}
\end{equation}

For fermionic part, the final form of the action 
derived in the conformal gauge is
\begin{equation}
S^{(2,c)}_{f,A} = - {1\over 2\pi}\int d\tau_c 
\int^{2\pi{E \over L^2}}_{0}
\! d\sigma_c \,
\theta^{T} 
\Big[\partial_{\tau_c} \theta
- i \ft{7-p}{4} \ell \bar{r}^{3-p \over 2} \gamma_{(p)}\theta
-i \gamma_{9}\partial_{\sigma_c}\theta\Big],
\label{eq:actionferconfa}
\end{equation}
\begin{equation}
S^{(2,c)}_{f,B} =  -{1\over 2\pi}\int d\tau_c 
\int^{2\pi{E \over L^2}}_{0}
\! d\sigma_c \,
(\theta^I)^{T} 
\Big[\partial_{\tau_c} \theta^I
- i \ft{7-p}{4} \ell \bar{r}^{3-p \over 2} \gamma_{(p)}\theta^I
+i 
s_2^{IJ}  \partial_{\sigma_c}\theta^J \Big].
\label{eq:actionferconfb}
\end{equation}
These actions are brought to the form (\ref{eq:SfA2h}) 
and (\ref{eq:SfB2h}) respectively, by the reparametrization (\ref{eq:wsredef}).

%%%%%%
\section{Relation between Poincar\'{e} and global
coordinates of AdS$_{p+2}$}
%%%
\setcounter{equation}{0}
\renewcommand{\theequation}{\Alph{section}.\arabic{equation}}
%%%%%%

We explain the relation between the  
Poincar\'{e} and global coordinates of AdS$_{p+2}$.
As mentioned in section~2, $(p+2)$-dimensional part 
(represented by $(t,r,x_a)$) of the near-horizon D$p$-brane metric 
is conformally equivalent to AdS$_{p+2}$.
Explicitly, the $(p+2)$-dimensional part of the metric  
is
\begin{eqnarray}
ds_{(p+2)}^2 &=& L^2 
\left( r^{7-p\over 2} (-dt^{2}+d x_{a}^{2})
+ r^{-{7-p \over 2}} dr^{2}  \right)
\\
&=& L^2 r^{-{3-p \over 2}}
\left(\ft{2}{5-p}\right)^{2}
ds_{{\rm AdS}_{p+2}}^2 
\end{eqnarray}
where 
$ds_{{\rm AdS}_{p+2}}^2$ is the metric of AdS$_{p+2}$ space: 
\begin{equation}
ds_{{\rm AdS}_{p+2}}^2 = {1\over z^2}(-dt^{2}+dx_{a}^{2}+ dz^2 ) ,\quad 
z=\ft{2}{5-p} r^{-{5-p\over 2}}.
\end{equation}

The Poincar\'{e} coordinates can be embedded into the flat 
${\rm R}^{p+1,2}$ as follows:
\begin{eqnarray}
X_0 &=& {z\over 2\tilde{\ell}}\left[ 1+{1\over z^2}
(\tilde{\ell}^2+x_a^2 -t^2)\right]
\,,\qquad
X_{p+2}  = %\tilde{\ell}\, 
{t\over z},
\nonumber\\
X_a &=& %\tilde{\ell}\, 
{x_a \over z}\,,\qquad
X_{p+1}  =  {z\over 2\tilde{\ell}}\left[ 
1-{1\over z^2}( \tilde{\ell}^2-x_a^2 +t^2 )
\right]
\label{eq:ctrans1}
\end{eqnarray}
where 
$$
X_0^2+X_{p+2}^2 -\sum_{i=1}^{p+1} X_i^2=1
 \, , \quad ds^{2}=-dX_{0}^{2}-dX_{p+2}^{2}+\sum_{i=1}^{p+1}dX_{i}^{2}
%\tilde{\ell}= \ft{2}{5-p}\ell
$$
By choosing the following parameterization 
\begin{equation}
X_0 = \cosh\!\rho\, \cos \!s
\,,\quad
X_i = \sinh\!\rho\, \Omega_i^{(p)}
\, ,\quad
X_{p+2} =  \cosh\!\rho\, \sin \!s,
\label{eq:ctrans2}
\end{equation}
we obtain the metric in global coordinates:
\begin{equation}
ds_{{\rm AdS}_{p+2}}^2 = 
d\rho^2+\sinh^2\!\rho \, (d\Omega_i^{(p)})^2 -\cosh^2\!\rho\, ds^2
\end{equation}
where $( d\Omega_i^{(p)})^2$ ($i=1,\cdots, p+1$) is
the metric on $S^{p}$ ($\sum_i (\Omega_i^{(p)})^2=1 $).
If we define
the following coordinates $\alpha_a$ ($a=1,\cdots, p$) on $S^p$:
\begin{equation}
\Omega_a^{(p)} = \left(\prod_{i=1}^{a-1} \sin\alpha_i \right) \cos\alpha_{a} ,\quad
\Omega_{p+1}^{(p)} = \prod_{i=1}^{p} \sin\alpha_i,
\end{equation}
$( d\Omega_i^{(p)})^2$ becomes
\begin{equation}
( d\Omega_i^{(p)})^2 = d\alpha_1^2 +\sin^2\!\alpha_1 d\alpha_2^2 +
\cdots + \left(\prod_{j=1}^{p-1} \sin^2\!\alpha_j \right)d\alpha_{p}^2.
\end{equation}

In the global coordinates, 
the RR field strength (\ref{eq:nhDp}) is written as
\begin{equation}
F_{p+2}= - L^{p+1}(7-p) \left(\ft{2}{5-p}\right)^{p+2}
r^{(3-p)^2 \over 2} \sinh^p\!\rho \cosh\!\rho \,ds\wedge dv^{p} \wedge d\rho
\label{eq:RRDpg}
\end{equation}
where $dv^{p}$ is the volume form of $S^{p}$ and 
$
r = \ell^{-{2\over 5-p}}
(\cosh\rho \, \cos s -\sinh\rho \, \Omega^{(p)}_{p+1})^{2\over 5-p}
.
$

For the background after the double Wick rotation (\ref{eq:metricDpdW}),
metric of the $(p+2)$-dimensional part is
\begin{equation}
d\tilde{s}_{(p+2)}^2 = L^2 r^{-{3-p \over 2}}
\left(\ft{2}{5-p}\right)^{2}
ds_{{\rm AdS}^E_{p+2}}^2 
\end{equation}
where 
$ds_{{\rm AdS}^E_{p+2}}^2$ is the metric of the Euclidean AdS$_{p+2}$ space 
(hyperbolic space) 
\begin{eqnarray}
ds_{{\rm AdS}^E_{p+2}}^2  & =& {1\over z^2}(dt^{2}+dx_{a}^{2}+ dz^2 )
\nonumber\\
&=&
d\rho^2+\sinh^2\!\rho \, (d\Omega_i^{(p)})^2 +\cosh^2\!\rho\, ds^2.
\end{eqnarray}
The relation between $(t,z,x_a)$ and $(s,\rho, \Omega_i^{(p)})$
is given by (\ref{eq:ctrans1}) and (\ref{eq:ctrans2}) with 
$X_{p+2}\to -i X_{p+2}$, $t\to -it$ and $s \to -is$.
In particular, we have
\begin{equation}
z =
\frac{\tilde{\ell} }
{\cosh\rho \, \cosh\!s -\sinh\rho \, \Omega^{(p)}_{p+1}}.
\end{equation}
We also see that the tunneling null geodesic 
$t^2+z^2=\tilde{\ell}^2$ is represented by 
$\rho=0$ in the global coordinate system.

%%%%%%
\section{$\kappa$-symmetry and supersymmetry 
in the semi-classical limit}
%%%
\setcounter{equation}{0}
\renewcommand{\theequation}{\Alph{section}.\arabic{equation}}
%%%%%%

We discuss the world-sheet supersymmetry of the superstring
action in the semi-classical limit $S^{(2)}=S^{(2)}_b+S^{(2)}_f$ 
given in section~2.
Here, we shall first take the Penrose limit of 
the double Wick rotated near-horizon D$p$-brane geometry,
and study Killing spinors of the background.
Superstring action in the light-cone gauge is equal
to the action for the fluctuation in the conformal gauge 
obtained in Appendix~A.

In the Penrose limit, the background (\ref{eq:metricDpdW}) becomes
\begin{equation}
  d\tilde{s}_p^2 = 2 dX^+ dX^- + dX_i^2 +dY_l^2 + 
(m_X({\scriptstyle X^+})^2 X_i^2 + m_Y({\scriptstyle X^+})^2 Y_l^2) {dX^+}^2 
\end{equation}
where $m_X({\scriptstyle X^+})^2$ and $m_Y({\scriptstyle X^+})^2$ 
are given by (\ref{eq:timedepMX}) and (\ref{eq:timedepMY}). 
The RR field strength correspondingly reduces to 
\begin{equation}
\tilde{F}_{p+2}= 
i \, (7\!-\!p ) \ell 
\bar{r}^{(3-p)(9-p) \over 4} \,dX^+ \wedge dX_1 \wedge \cdots \wedge dX_{p+1} .
\end{equation}
Here $\bar{r}=\bar{r}({\scriptstyle X^+})$ is defined by 
${d\bar{r}/dX^+}=\pm \sqrt{\ell^2 \bar{r}^{5-p}-1}$ as in (\ref{eq:defrtau}).
The Green-Schwarz action on this background in the light-cone gauge
\begin{equation}
\sqrt{h}h^{\alpha\beta} =\delta^{\alpha\beta},\quad 
X^+ = \tau_c, \quad \Gamma_- \Theta=0,   
\end{equation}
is equal to the action
$S^{(2,c)}_b + S^{(2,c)}_f$ in (\ref{eq:actionbosconf}),
 (\ref{eq:actionferconfa}), (\ref{eq:actionferconfb}).

The action is invariant under the local 
$\kappa$-symmetry and the space-time supersymmetry
when the background preserves supersymmetry.
The transformation law is~\cite{CvLuPoSt}
\begin{equation}
\delta\Theta =(1+\Gamma)\kappa +\epsilon ,\quad
\delta X^{\mu}= -i \bar{\Theta} \Gamma^{\mu}(1+\Gamma)\kappa
+i \bar{\Theta} \Gamma^\mu \epsilon
\end{equation}
with $\Gamma ={i\over 2 \sqrt{h}} \epsilon^{\alpha\beta}
\partial_{\alpha}X^{\mu} \partial_{\beta}X^{\nu} \Gamma_{\mu\nu}\Gamma_{11} $
for type IIA, and 
\begin{equation}
\delta\Theta^I =(\delta^{IJ}+\Gamma^{IJ})\kappa^J +\epsilon^I ,\quad
\delta X^{\mu}= i \bar{\Theta}^I \Gamma^{\mu}s_0^{IJ} (1+\Gamma)\kappa^J
-i \bar{\Theta}^{I} \Gamma^\mu s_0^{IJ} \epsilon^J
\label{eq:dthdxa}
\end{equation}
with  
$\Gamma^{IJ} ={i\over 2 \sqrt{h}} \epsilon^{\alpha\beta}
\partial_{\alpha}X^{\mu} \partial_{\beta}X^{\nu} \Gamma_{\mu\nu}s_2^{IJ}$
for type IIB. Here 
$\epsilon$ is a Killing spinor of the background
which satisfies
\begin{equation}
{\cal D}_\mu\epsilon = (\nabla_{\mu} + \Omega \Gamma_\mu)\epsilon=0.
\label{eq:dthdxb}
\end{equation}

Light-cone gauge condition $\Gamma_{-}\Theta=0$ must be
kept under the transformation.
For Killing spinors satisfying $\Gamma_{-}\epsilon=0$, 
the transformation (\ref{eq:dthdxa}) or (\ref{eq:dthdxb}) 
with $\kappa=0$ is just the trivial fermionic shift symmetry 
$\delta\theta \sim \epsilon$, $\delta X^{A}=0$ ($A=1,\ldots, 8$).
When there are Killing spinors such that $\Gamma_{-}\epsilon\ne 0$,
world-sheet supersymmetry is realized by 
the transformation $\delta(\Gamma_{+}\Theta)$, $\delta X^{A}$,
where  $\kappa$ is chosen such that $\delta(\Gamma_{-}\Theta)=0$.

For $p\ne 3$, we have only 16 Killing spinors with
$\Gamma_{-}\epsilon=0$, which exist in the general pp-wave backgrounds,
and there is no world-sheet supersymmetry. 
On the other hand, for $p=3$, the background has 
32 Killing spinors, and the extra 16 supersymmetries with
$\Gamma_{-}\epsilon\ne 0$ generates world-sheet supersymmetry.

%We must adjust the $\kappa$-symmetry parameter $\kappa$ 
%in order to keep the light-cone gauge
%$\delta (\Gamma_- \Theta)=0$. 
%%keep the gauge condition $\Gamma_{-}\Theta=0$.
%The residual symmetry $\delta (\Gamma_{+} \Theta)$ 
%is realized as world-sheet symmetry~\cite{CvLuPoSt}.
%
%For $p\ne 3$, we have only 16 Killing spinors, 
%which satisfy $\Gamma_{-}\epsilon=0$
%and generally exist in the pp-wave background. 
%These generate the trivial fermionic shift symmetries 
% 
%but there is no linearly realized world-sheet supersymmetry. 
%On the other hand, for $p=3$, the background has 
%32 Killing spinors, and the extra 16 supersymmetries
%realizes the world-sheet supersymmetry.

%%%%%%%%%%%%%%%%%%%%%%%%%%%%%%%%%%%%
%%%%%%%%%%%%%%%%%%%%%%%%%%%%%%%%%%%%

\end{document}